\documentclass{article}%
\usepackage{amsmath}
\usepackage{graphicx}%
\usepackage{amsfonts}%
\usepackage{amssymb}
\newtheorem{theorem}{Theorem}[section]

\newtheorem{definition}[theorem]{Definition}
\newtheorem{example}[theorem]{Example}

\newtheorem{notation}[theorem]{Notation}

\newtheorem{remark}[theorem]{Remark}

\newenvironment{proof}[1][Proof]{\textbf{#1.} }{\ \rule{0.5em}{0.5em}}

\begin{document}

\title{Asynchronous pseudo-systems}
\author{Serban. E. Vlad\\Oradea City Hall, Piata Unirii, Nr. 1, 410100, Oradea, Romania\\serban\_e\_vlad@yahoo.com, http://www.geocities.com/serban\_e\_vlad}
\date{}
\maketitle

\begin{abstract}
The paper introduces the concept of asynchronous pseudo-system. Its purpose is
to correct/generalize/continue the study of the asynchronous systems (the
models of the asynchronous circuits) that has been started in [1], [2].

\end{abstract}

$\bigskip$

\textbf{\bigskip AMS Classification}: 93A10

\bigskip\textbf{Keywords}: signal, asynchronous pseudo-system, asynchronous system\bigskip

\textbf{Contents }1.\textbf{ }Introduction, 2. Differentiable functions.
Signals, 3. Pseudo-systems, 4. Initial states and final states, 5. Initial
time and final time, 6. Initial state function and final state function, 7.
Pseudo-subsystems, 8. Dual pseudo-systems, 9. Inverse pseudo-systems, 10.
Direct product, 11. Parallel connection, 12. Serial connection, 13.
Complement, 14. Intersection and reunion, 15. Systems

\section{Introduction}

The study of the asynchronous systems [1], [2] was generated by the
mathematical models of the asynchronous circuits from the digital electrical
engineering. What we have proposed there is that starting from the very
general notion of system (non-deterministic, in the input-output sense), by
the addition of definitions=axioms to rediscover one by one the properties of
the models of the asynchronous circuits. Roughly speaking, the signals are the
differentiable, right continuous $\mathbf{R}\rightarrow\{0,1\}^{n}$ functions
with initial values (i.e. with limit when $t\rightarrow-\infty$) and an
(asynchronous) system is a multi-valued function that associates to a signal
$\mathbf{R}\rightarrow\{0,1\}^{m}$ called (admissible) input, a non-empty set
of $\mathbf{R}\rightarrow\{0,1\}^{n}$ signals, called (possible) states.

The purpose of this work is that of correcting/improving/generalizing the
frame of these papers and the main concept is that of pseudo-system,
representing a multi-valued function from differentiable right continuous
$\mathbf{R}\rightarrow\{0,1\}^{m}$ functions called inputs to (empty or
non-empty) sets of differentiable right continuous $\mathbf{R}\rightarrow
\{0,1\}^{n}$ functions, called states. In other words, we have relaxed two
conditions relative to the systems:

- the functions $\mathbf{R}\rightarrow\{0,1\}^{n}$ without limit when
$t\rightarrow-\infty$ (without initial values) are accepted

- to an input $u:\mathbf{R}\rightarrow\{0,1\}^{m}$ there may correspond an
empty set of states, i.e. we accept the existence of non-admissible inputs.

We prefer this approach in order to underline the duality between the initial
states and initial time on one hand and the final states and final time, on
the other hand. Besides, we must take into account the fact that very simple
circuits like the RS latch for example have non-admissible inputs ($R\cdot
S=1$).

We define and characterize the pseudo-systems, the initial \ and the final
states, the initial and the final time, the initial and the final state
functions, the pseudo-subsystems, the dual pseudo-systems, the inverse
pseudo-systems, the direct product, the parallel and the serial connection,
the complement, the intersection and the reunion of the pseudo-systems. The
conclusions are expressed in the last section, where we define the systems as
special cases of pseudo-systems whose admissible inputs and possible states
are signals and we also show how the previous topics related with the
pseudo-systems are particularized to the case of the systems.

We have written in full details all the dual results. The proofs are generally
elementary and some of them have been omitted, some of them have been included
for the reason of making the exposure as readable as possible. The dual proofs
have been omitted.

\section{Differentiable functions. Signals}

We note with $\mathbf{B}=\{0,1\}$ the Boole algebra with two elements and with
$\chi_{A}:\mathbf{R}\rightarrow\mathbf{B}$ the characteristic function of the
set $A\subset\mathbf{R}$. The differentiable functions $x:\mathbf{R}%
\rightarrow\mathbf{B}^{n}$ are by definition of the form:%
\begin{equation}
x(t)=...\oplus x(t_{-1})\cdot\chi_{\{t_{-1}\}}(t)\oplus x(%
\genfrac{.}{.}{}{}{t_{-1}+t_{0}}{2}%
)\cdot\chi_{(t_{-1},t_{0})}(t)\oplus\label{e0}%
\end{equation}%
\[
\oplus x(t_{0})\cdot\chi_{\{t_{0}\}}(t)\oplus x(%
\genfrac{.}{.}{}{}{t_{0}+t_{1}}{2}%
)\cdot\chi_{(t_{0},t_{1})}(t)\oplus x(t_{1})\cdot\chi_{\{t_{1}\}}(t)\oplus...
\]
where $...<t_{-1}<t_{0}<t_{1}<...$ is an upper and lower unbounded sequence
and $\mathbf{R}$ is the dense ($\forall t\in\mathbf{R},\forall t^{\prime}%
\in\mathbf{R},t<t^{\prime}\Longrightarrow\exists t"\in\mathbf{R}%
,t<t"<t^{\prime}$) and linear (i.e. totally ordered: $\forall t\in
\mathbf{R},\forall t^{\prime}\in\mathbf{R},t\leq t^{\prime}\quad or\quad
t^{\prime}\leq t$) time set. If in (\ref{e0}) $x(t_{k})=x(%
\genfrac{.}{.}{}{}{t_{k}+t_{k+1}}{2}%
),k\in\mathbf{Z}$, then $x$ is right continuous and it is of the form
\[
x(t)=...\oplus x(t_{-1})\cdot\chi_{\lbrack t_{-1},t_{0})}(t)\oplus
x(t_{0})\cdot\chi_{\lbrack t_{0},t_{1})}(t)\oplus...
\]
The set of the ($n$-dimensional) differentiable, right continuous functions
$x$ is noted with $\widetilde{S}^{(n)}.\footnote{The differentiable left
continuous functions%
\[
x(t)=...\oplus x(t_{0})\cdot\chi_{(t_{-1},t_{0}]}(t)\oplus x(t_{1})\cdot
\chi_{(t_{0},t_{1}]}(t)\oplus...
\]
give an equivalent manner of writing this paper. In previous works we have
associated non-anticipation with right continuity and anticipation with left
continuity.}$

We consider the next properties of some $x\in\widetilde{S}^{(n)}:$%
\begin{equation}
\exists\mu\in\mathbf{B}^{n},\exists t_{0}\in\mathbf{R},\forall t<t_{0}%
,x(t)=\mu\label{e1}%
\end{equation}%
\begin{equation}
\exists\mu^{\prime}\in\mathbf{B}^{n},\exists t_{f}\in\mathbf{R},\forall
t>t_{f},x(t)=\mu^{\prime} \label{e3}%
\end{equation}
where in (\ref{e1}) $\exists\mu\in\mathbf{B}^{n},\exists t_{0}\in\mathbf{R}$
commute and in (\ref{e3}) $\exists\mu^{\prime}\in\mathbf{B}^{n},\exists
t_{f}\in\mathbf{R}$ commute also.

If (\ref{e1}) is fulfilled:

- $\mu$ is unique and is called the initial value of $x$. We shall note it
sometimes with $\underset{t\rightarrow-\infty}{\lim}x(t)$, $x(-\infty+0)$ or
with $x(t_{0}-0)$

- $t_{0}$ is not unique, since any $t_{0}^{\prime}<t_{0}$ satisfies (\ref{e1})
too. It is called the initial time of $x$.

If (\ref{e3}) is fulfilled:

- $\mu^{\prime}$ is unique and is called the final value of $x$. The usual
notations are $\underset{t\rightarrow\infty}{\lim}x(t)$, $x(t_{f})$ and
$x(\infty-0)$

- $t_{f}$ is not unique, because any $t_{f}^{\prime}>t_{f}$ satisfies
(\ref{e3}) too. It is called the final time of $x$.

We call ($n-$dimensional) signal a function $x\in\widetilde{S}^{(n)}$ with the
property that (\ref{e1}) is satisfied. The signals are represented under the
form:%
\[
x(t)=x(t_{0}-0)\cdot\chi_{(-\infty,t_{0})}(t)\oplus x(t_{0})\cdot\chi_{\lbrack
t_{0},t_{1})}(t)\oplus x(t_{1})\cdot\chi_{\lbrack t_{1},t_{2})}(t)\oplus...
\]
where $t_{0}<t_{1}<t_{2}<...$ is unbounded. The set of the signals is noted
with $S^{(n)}$.

Dually, we call ($n-$dimensional) signal$^{\ast}$ a function $x\in
\widetilde{S}^{(n)}$ with the property that (\ref{e3}) is true and such
functions are represented under the form%
\[
x(t)=...\oplus x(t_{-2})\cdot\chi_{\lbrack t_{-2},t_{-1})}(t)\oplus
x(t_{-1})\cdot\chi_{\lbrack t_{-1},t_{0})}(t)\oplus x(t_{0})\cdot\chi_{\lbrack
t_{0},\infty)}(t)
\]
where $...<t_{-2}<t_{-1}<t_{0}$ is unbounded. The set of the signals$^{\ast}$
is noted with $S^{(n)\ast}$.

We shall often write $\widetilde{S},$ $S,$ $S^{\ast}$ instead of
$\widetilde{S}^{(1)},$ $S^{(1)},$ $S^{(1)\ast}$.

We use the notations $P(L)=\{K|K\subset L\}$ and $P^{\ast}(L)=\{K|K\subset
L,K\neq\emptyset\}$, where $L$ is any of $\mathbf{B}^{n},$ $\widetilde
{S}^{(n)},$ $S^{(n)},$ $S_{{}}^{(n)\ast}$.

\section{Pseudo-systems}

\begin{definition}
The functions $f:\widetilde{S}^{(m)}\rightarrow P(\widetilde{S}^{(n)})$ are
called (asynchronous) pseudo-systems. The elements $u\in\widetilde{S}^{(m)}$
are called inputs (in the pseudo-system): admissible if $f(u)\neq\emptyset$
and non-admissible otherwise, while the elements $x\in f(u)$ are called
(possible) states (of the pseudo-system), or (possible) outputs (from the pseudo-system).
\end{definition}

\begin{remark}
The pseudo-systems are multi-valued functions that associate to each input $u$
the set of the possible states $f(u)$, the origin of the concept being
situated in the modeling of the asynchronous circuits from digital engineering.

A non-admissible input, i.e. an input for which $f(u)=\emptyset$, is
considered a cause of no effect that can be expressed by $f$ and the null
pseudo-system that is defined by $\forall u\in\widetilde{S}^{(m)}%
,f(u)=\emptyset$ represents the limit situation when $f$ does not express a
determination between the elements of $\widetilde{S}^{(m)}$ and the elements
of $\widetilde{S}^{(n)}$ ($f$ models nothing). The other limit situation is
represented by the total pseudo-system, defined by $\forall u\in\widetilde
{S}^{(m)},f(u)=\widetilde{S}^{(n)}$ ($f$ models every circuit with
$m$-dimensional inputs and $n$-dimensional outputs); for this pseudo-system,
all the inputs are admissible.

The multi-valued character of the cause-effect association is due to
statistical fluctuations in the fabrication process, variations in ambient
temperature, power supply etc.

In applications, the pseudo-systems are defined sometimes not explicitly, like
before, but implicitly, by a system of equations and inequalities where $u$ is
given, $t$ is the time variable, $x$ is the unknown and the temporal logical
connectors depending on them are differentiable in general (they are not right continuous).
\end{remark}

\begin{example}
The pseudo-system $f:\widetilde{S}\rightarrow P(\widetilde{S})$ is defined by
the double inequality%
\begin{equation}
\underset{\xi\in\lbrack t-d,t)}{\bigcap}u(\xi)\leq x(t)\leq\underset{\xi
\in\lbrack t-d,t)}{\bigcup}u(\xi) \label{e6.1}%
\end{equation}
where $d>0$. When $u,x\in\widetilde{S}$, the connectors $\underset{\xi
\in\lbrack t-d,t)}{\bigcap}u(\xi)$ and $\underset{\xi\in\lbrack t-d,t)}%
{\bigcup}u(\xi)$ are just differentiable, they are not right continuous.
\end{example}

\section{Initial states and final states}

\begin{remark}
We state the next properties of the pseudo-system $f$:%
\begin{equation}
\forall u\in\widetilde{S}^{(m)},\forall x\in f(u),\exists\mu\in\mathbf{B}%
^{n},\exists t_{0}\in\mathbf{R},\forall t<t_{0},x(t)=\mu\label{e2.1}%
\end{equation}%
\begin{equation}
\forall u\in\widetilde{S}^{(m)},\exists\mu\in\mathbf{B}^{n},\forall x\in
f(u),\exists t_{0}\in\mathbf{R},\forall t<t_{0},x(t)=\mu\label{e2.2}%
\end{equation}%
\begin{equation}
\exists\mu\in\mathbf{B}^{n},\forall u\in\widetilde{S}^{(m)},\forall x\in
f(u),\exists t_{0}\in\mathbf{R},\forall t<t_{0},x(t)=\mu\label{e2.3}%
\end{equation}%
\begin{equation}
\forall u\in\widetilde{S}^{(m)},\forall x\in f(u),\exists\mu\in\mathbf{B}%
^{n},\exists t_{f}\in\mathbf{R},\forall t>t_{f},x(t)=\mu\label{e2.4}%
\end{equation}%
\begin{equation}
\forall u\in\widetilde{S}^{(m)},\exists\mu\in\mathbf{B}^{n},\forall x\in
f(u),\exists t_{f}\in\mathbf{R},\forall t>t_{f},x(t)=\mu\label{e2.5}%
\end{equation}%
\begin{equation}
\exists\mu\in\mathbf{B}^{n},\forall u\in\widetilde{S}^{(m)},\forall x\in
f(u),\exists t_{f}\in\mathbf{R},\forall t>t_{f},x(t)=\mu\label{e2.6}%
\end{equation}
where in (\ref{e2.1}) $\mu$ and $t_{0}$ depend on $x$ only, thus $\exists
\mu\in\mathbf{B}^{n},\exists t_{0}\in\mathbf{R}$ commute and similarly for
$\mu$ and $t_{f}$ in (\ref{e2.4}). We observe the dualities between
(\ref{e2.1}) and (\ref{e2.4}); (\ref{e2.2}) and (\ref{e2.5}); (\ref{e2.3}) and
(\ref{e2.6}) and on the other hand we remark the truth of the implications%
\[
(\ref{e2.3})\Longrightarrow(\ref{e2.2})\Longrightarrow(\ref{e2.1})
\]%
\[
(\ref{e2.6})\Longrightarrow(\ref{e2.5})\Longrightarrow(\ref{e2.4})
\]
If $f$ is the null pseudo-system, it fulfills trivially all the properties
(\ref{e2.1}),...,(\ref{e2.6}). If (\ref{e2.1}) is true with $f$ non-null, it
defines a partial function $\widetilde{S}^{(n)}\rightarrow\mathbf{B}^{n}$ that
associates to each $x\in\underset{u\in\widetilde{S}^{(m)}}{\bigcup}f(u)$ its
initial value $\mu$. If (\ref{e2.2}) is true with $f$ non-null, it defines a
partial function $\widetilde{S}^{(m)}\rightarrow\mathbf{B}^{n}$ that
associates to each admissible input $u$ the common initial value $\mu$ of all
$x\in f(u)$. Dually, if $f$ is non-null and (\ref{e2.4}), (\ref{e2.5}) are
true, they define two partial functions $\widetilde{S}^{(n)}\rightarrow
\mathbf{B}^{n}$ and $\widetilde{S}^{(m)}\rightarrow\mathbf{B}^{n}$.

If $f$ is null, any $\mu\in\mathbf{B}^{n}$ makes (\ref{e2.3}) and (\ref{e2.6})
true; otherwise, the value of $\mu$ is uniquely defined by either of
(\ref{e2.3}) and (\ref{e2.6}).
\end{remark}

\begin{definition}
If $f$ satisfies (\ref{e2.1}), we say that it has initial states. The vectors
$\mu$ are called in this case (the) initial states (of $f$), or (the) initial
values of the states (of $f$).
\end{definition}

\begin{definition}
We suppose that $f$ satisfies (\ref{e2.2}). We say in this situation that it
has race-free, or delay-insensitive initial states and the initial states
$\mu$ are called race-free, or delay-insensitive themselves.
\end{definition}

\begin{definition}
When $f$ satisfies (\ref{e2.3}), we use to say that it has a (constant)
initial state $\mu$. We say in this case that $f$ is initialized and that
$\mu$ is its (constant) initial state.
\end{definition}

\begin{definition}
If $f$ satisfies (\ref{e2.4}), it is called absolutely stable and we also say
that it has final states. The vectors $\mu$ have in this case the name of
final states (of $f$), or of final values of the states (of $f$), or of steady
states (of $f$), or of steady values of the states (of $f$).
\end{definition}

\begin{definition}
If $f$ fulfills the property (\ref{e2.5}), it is called absolutely race-free
stable, or absolutely delay-insensitive and we also say that it has race-free
final states. The final states $\mu$ are called in this case race-free, or delay-insensitive.
\end{definition}

\begin{definition}
We suppose that the pseudo-system $f$ satisfies (\ref{e2.6}). Then it is
called absolutely constantly stable or equivalently we say that it has a
(constant) final state. The vector $\mu$ is called in this situation
(constant) final state.
\end{definition}

\begin{remark}
The previous terminology is related with the dualities initial-final,
initialized-absolutely stable as well as with hardware engineering. In
hardware engineering, 'race' means: 'which coordinate of $x$ switches first is
the winner' or perhaps 'several ways to go' and in this case 'race-free' means
'one way to go'; and delay-insensitivity means (vaguely) 'for any fluctuations
in the fabrication process', see Remark 3.2.
\end{remark}

\section{Initial time and final time}

\begin{remark}
We state the next properties on the pseudo-system $f$:%
\begin{equation}
\forall u\in\widetilde{S}^{(m)},\forall x\in f(u)\cap S_{{}}^{(n)},\exists
\mu\in\mathbf{B}^{n},\exists t_{0}\in\mathbf{R},\forall t<t_{0},x(t)=\mu
\label{e3.1}%
\end{equation}%
\begin{equation}
\forall u\in\widetilde{S}^{(m)},\exists t_{0}\in\mathbf{R},\forall x\in
f(u)\cap S_{{}}^{(n)},\exists\mu\in\mathbf{B}^{n},\forall t<t_{0}%
,x(t)=\mu\label{e3.2}%
\end{equation}%
\begin{equation}
\exists t_{0}\in\mathbf{R},\forall u\in\widetilde{S}^{(m)},\forall x\in
f(u)\cap S_{{}}^{(n)},\exists\mu\in\mathbf{B}^{n},\forall t<t_{0}%
,x(t)=\mu\label{e3.3}%
\end{equation}%
\begin{equation}
\forall u\in\widetilde{S}^{(m)},\forall x\in f(u)\cap S_{{}}^{(n)\ast}%
,\exists\mu\in\mathbf{B}^{n},\exists t_{f}\in\mathbf{R},\forall t>t_{f}%
,x(t)=\mu\label{e3.4}%
\end{equation}%
\begin{equation}
\forall u\in\widetilde{S}^{(m)},\exists t_{f}\in\mathbf{R},\forall x\in
f(u)\cap S_{{}}^{(n)\ast},\exists\mu\in\mathbf{B}^{n},\forall t>t_{f}%
,x(t)=\mu\label{e3.5}%
\end{equation}%
\begin{equation}
\exists t_{f}\in\mathbf{R},\forall u\in\widetilde{S}^{(m)},\forall x\in
f(u)\cap S_{{}}^{(n)\ast},\exists\mu\in\mathbf{B}^{n},\forall t>t_{f}%
,x(t)=\mu\label{e3.6}%
\end{equation}
where in (\ref{e3.1}) $\mu$ and $t_{0}$ depend on $x$ only, making $\exists
\mu\in\mathbf{B}^{n},\exists t_{0}\in\mathbf{R}$ commute and the situation is
similar for $\mu$ and $t_{f}$ in (\ref{e3.4}).

The properties (\ref{e3.1}) and (\ref{e3.4}) are fulfilled by all the
pseudo-systems and they are present here for the symmetry of the exposure only.

The dualities between (\ref{e3.1}) and (\ref{e3.4}); (\ref{e3.2}) and
(\ref{e3.5}); (\ref{e3.3}) and (\ref{e3.6}) take place and the next
implications hold:%
\[
(\ref{e3.3})\Longrightarrow(\ref{e3.2})\Longrightarrow(\ref{e3.1})
\]%
\[
(\ref{e3.6})\Longrightarrow(\ref{e3.5})\Longrightarrow(\ref{e3.4})
\]
If $f$ is the null pseudo-system or, more generally, if in one of
(\ref{e3.1}),...,(\ref{e3.3}) $\forall u\in\widetilde{S}^{(m)},$ $f(u)\cap
S_{{}}^{(n)}=\emptyset$, that property is trivially fulfilled. Here the
similarity with Remark 4.1 ends, since defining a partial function
$\widetilde{S}^{(n)}\rightarrow\mathbf{R}$ for example in the case of
(\ref{e3.1}) associating to each state $x\in f(u)\cap S_{{}}^{(n)}$ its
initial time is not quite natural. Reasoning is the same for the final time.
\end{remark}

\begin{definition}
If $f$ satisfies (\ref{e3.1}), we say that it has unbounded initial time and
any $t_{0}$ satisfying this property is called unbounded initial time (instant).
\end{definition}

\begin{definition}
Let $f$ fulfilling the property (\ref{e3.2}). We say that it has bounded
initial time and any $t_{0}$ making this property true is called bounded
initial time (instant).
\end{definition}

\begin{definition}
When $f$ satisfies (\ref{e3.3}), we use to say that it has fix, or universal
initial time and any $t_{0}$ fulfilling (\ref{e3.3}) is called fix (or
universal) initial time (instant).
\end{definition}

\begin{definition}
We suppose that $f$ satisfies (\ref{e3.4}). Then we say that it has unbounded
final time and any $t_{f}$ satisfying this property is called unbounded final
time (instant).
\end{definition}

\begin{definition}
If $f$ fulfills the property (\ref{e3.5}), we say that it has bounded final
time. Any number $t_{f}$ satisfying (\ref{e3.5}) is called bounded final time (instant).
\end{definition}

\begin{definition}
We suppose that the pseudo-system $f$ satisfies the property (\ref{e3.6}).
Then we say that it has fix, or universal final time and any number $t_{f}$
satisfying (\ref{e3.6}) is called fix, or universal final time (instant).
\end{definition}

\begin{theorem}
If the pseudo-system $f$ has initial states, then the next non-exclusive
possibilities exist:

a) $f$ has initial states and unbounded initial time%
\[
\forall u\in\widetilde{S}^{(m)},\forall x\in f(u),\exists\mu\in\mathbf{B}%
^{n},\exists t_{0}\in\mathbf{R},\forall t<t_{0},x(t)=\mu
\]
where $\mu$ and $t_{0}$ depend on $x$ only, thus $\exists\mu\in\mathbf{B}%
^{n},\exists t_{0}\in\mathbf{R}$ commute

b) $f$ has initial states and bounded initial time%
\[
\forall u\in\widetilde{S}^{(m)},\exists t_{0}\in\mathbf{R},\forall x\in
f(u),\exists\mu\in\mathbf{B}^{n},\forall t<t_{0},x(t)=\mu
\]

c) $f$ has initial states and fix initial time%
\[
\exists t_{0}\in\mathbf{R},\forall u\in\widetilde{S}^{(m)},\forall x\in
f(u),\exists\mu\in\mathbf{B}^{n},\forall t<t_{0},x(t)=\mu
\]

d) $f$ has race-free initial states and unbounded initial time%
\[
\forall u\in\widetilde{S}^{(m)},\exists\mu\in\mathbf{B}^{n},\forall x\in
f(u),\exists t_{0}\in\mathbf{R},\forall t<t_{0},x(t)=\mu
\]

e) $f$ has race-free initial states and bounded initial time%
\[
\forall u\in\widetilde{S}^{(m)},\exists\mu\in\mathbf{B}^{n},\exists t_{0}%
\in\mathbf{R},\forall x\in f(u),\forall t<t_{0},x(t)=\mu
\]
where $\mu$ and $t_{0}$ depend on $u$ only, thus $\exists\mu\in\mathbf{B}%
^{n},\exists t_{0}\in\mathbf{R}$ commute

f) $f$ has race-free initial states and fix initial time%
\[
\exists t_{0}\in\mathbf{R},\forall u\in\widetilde{S}^{(m)},\exists\mu
\in\mathbf{B}^{n},\forall x\in f(u),\forall t<t_{0},x(t)=\mu
\]

g) $f$ has a constant initial state and unbounded initial time%
\[
\exists\mu\in\mathbf{B}^{n},\forall u\in\widetilde{S}^{(m)},\forall x\in
f(u),\exists t_{0}\in\mathbf{R},\forall t<t_{0},x(t)=\mu
\]

h) $f$ has a constant initial state and bounded initial time%
\[
\exists\mu\in\mathbf{B}^{n},\forall u\in\widetilde{S}^{(m)},\exists t_{0}%
\in\mathbf{R},\forall x\in f(u),\forall t<t_{0},x(t)=\mu
\]

i) $f$ has a constant initial state and fix initial time%
\[
\exists\mu\in\mathbf{B}^{n},\exists t_{0}\in\mathbf{R},\forall u\in
\widetilde{S}^{(m)},\forall x\in f(u),\forall t<t_{0},x(t)=\mu
\]
where $\exists\mu\in\mathbf{B}^{n},\exists t_{0}\in\mathbf{R}$ commute.
\end{theorem}

\begin{proof}
e) We must show that the conjunction of (\ref{e2.2}) and (\ref{e3.2}) on one
hand and%
\begin{equation}
\forall u\in\widetilde{S}^{(m)},\exists\mu\in\mathbf{B}^{n},\exists t_{0}%
\in\mathbf{R},\forall x\in f(u),\forall t<t_{0},x(t)=\mu\label{e3.7}%
\end{equation}
where $\mu$ and $t_{0}$ depend on $u$ only (making $\exists\mu\in
\mathbf{B}^{n},\exists t_{0}\in\mathbf{R}$ commute) on the other hand - are
equivalent. This fact is obvious if $f$ is null, thus we can suppose that $f$
is non null and it is sufficient to consider some admissible arbitrary fixed
$u\in\widetilde{S}^{(m)}$.

(\ref{e2.2}) and (\ref{e3.2}) $\Longrightarrow$ (\ref{e3.7})

From (\ref{e2.2}) we have the existence of a unique $\mu\in\mathbf{B}^{n}$
depending on $u$ so that $\forall x\in f(u),x(-\infty+0)=\mu$ from where
$f(u)\subset S_{{}}^{(n)}$ and $f(u)\cap S_{{}}^{(n)}=f(u)$. From (\ref{e3.2})
we infer that%
\[
\exists t_{0}\in\mathbf{R},\forall x\in f(u),\forall t<t_{0},x(t)=\mu
\]
where $t_{0}$ depends on $u$ and the statements%
\[
\exists t_{0}\in\mathbf{R},\exists\mu\in\mathbf{B}^{n},\forall x\in
f(u),\forall t<t_{0},x(t)=\mu
\]%
\[
\exists\mu\in\mathbf{B}^{n},\exists t_{0}\in\mathbf{R},\forall x\in
f(u),\forall t<t_{0},x(t)=\mu
\]
are both true, as $\mu$ and $t_{0}$ depend on $u$ only. (\ref{e3.7}) is true.

(\ref{e3.7}) $\Longrightarrow$ (\ref{e2.2}) and (\ref{e3.2})

(\ref{e3.7}) $\Longrightarrow$ (\ref{e2.2}) is obvious. On the other hand a
unique $\mu\in\mathbf{B}^{n}$ exists so that%
\[
\exists t_{0}\in\mathbf{R},\forall x\in f(u),\forall t<t_{0},x(t)=\mu
\]
in particular the statement%
\[
\exists t_{0}\in\mathbf{R},\forall x\in f(u)\cap S_{{}}^{(n)},\forall
t<t_{0},x(t)=\mu
\]
is true, as well as%
\[
\exists t_{0}\in\mathbf{R},\forall x\in f(u)\cap S_{{}}^{(n)},\exists\mu
\in\mathbf{B}^{n},\forall t<t_{0},x(t)=\mu
\]
i.e. (\ref{e3.2}).
\end{proof}

\begin{theorem}
The next non-exclusive possibilities exist for the absolutely stable
pseudo-system $f$:

a) $f$ is absolutely stable with unbounded final time:%
\[
\forall u\in\widetilde{S}^{(m)},\forall x\in f(u),\exists\mu\in\mathbf{B}%
^{n},\exists t_{f}\in\mathbf{R},\forall t>t_{f},x(t)=\mu
\]
where $\mu$ and $t_{f}$ depend on $x$ only, thus $\exists\mu\in\mathbf{B}%
^{n},\exists t_{f}\in\mathbf{R}$ commute

b) $f$ is absolutely stable with bounded final time:%
\[
\forall u\in\widetilde{S}^{(m)},\exists t_{f}\in\mathbf{R},\forall x\in
f(u),\exists\mu\in\mathbf{B}^{n},\forall t>t_{f},x(t)=\mu
\]

c) $f$ is absolutely stable with fix final time:%
\[
\exists t_{f}\in\mathbf{R},\forall u\in\widetilde{S}^{(m)},\forall x\in
f(u),\exists\mu\in\mathbf{B}^{n},\forall t>t_{f},x(t)=\mu
\]

d) $f$ is absolutely race-free stable with unbounded final time:%
\[
\forall u\in\widetilde{S}^{(m)},\exists\mu\in\mathbf{B}^{n},\forall x\in
f(u),\exists t_{f}\in\mathbf{R},\forall t>t_{f},x(t)=\mu
\]

e) $f$ is absolutely race-free stable with bounded final time:%
\[
\forall u\in\widetilde{S}^{(m)},\exists\mu\in\mathbf{B}^{n},\exists t_{f}%
\in\mathbf{R},\forall x\in f(u),\forall t>t_{f},x(t)=\mu
\]
where $\mu$ and $t_{f}$ depend on $u$ only, thus $\exists\mu\in\mathbf{B}%
^{n},\exists t_{f}\in\mathbf{R}$ commute

f) $f$ is absolutely race-free stable with fix final time:%
\[
\exists t_{f}\in\mathbf{R},\forall u\in\widetilde{S}^{(m)},\exists\mu
\in\mathbf{B}^{n},\forall x\in f(u),\forall t>t_{f},x(t)=\mu
\]

g) $f$ is absolutely constantly stable with unbounded final time:%
\[
\exists\mu\in\mathbf{B}^{n},\forall u\in\widetilde{S}^{(m)},\forall x\in
f(u),\exists t_{f}\in\mathbf{R},\forall t>t_{f},x(t)=\mu
\]

h) $f$ is absolutely constantly stable with bounded final time:%
\[
\exists\mu\in\mathbf{B}^{n},\forall u\in\widetilde{S}^{(m)},\exists t_{f}%
\in\mathbf{R},\forall x\in f(u),\forall t>t_{f},x(t)=\mu
\]

i) $f$ is absolutely constantly stable with fix final time:%
\[
\exists\mu\in\mathbf{B}^{n},\exists t_{f}\in\mathbf{R},\forall u\in
\widetilde{S}^{(m)},\forall x\in f(u),\forall t>t_{f},x(t)=\mu
\]
where $\mu$ and $t_{f}$ are independent on each other, thus $\exists\mu
\in\mathbf{B}^{n},\exists t_{f}\in\mathbf{R}$ commute.
\end{theorem}

\begin{remark}
All the pseudo-systems have unbounded initial (final) time, the problem is if
they have initial (final) states or not. On the other hand, at both previous
theorems, the next implications hold:%
\[%
\begin{array}
[c]{ccccc}%
i) & \Longrightarrow & h) & \Longrightarrow & g)\\
\Downarrow &  & \Downarrow &  & \Downarrow\\
f) & \Longrightarrow & e) & \Longrightarrow & d)\\
\Downarrow &  & \Downarrow &  & \Downarrow\\
c) & \Longrightarrow & b) & \Longrightarrow & a)
\end{array}
\]
\end{remark}

\section{Initial state function and final state function}

\begin{definition}
Let the pseudo-system $f:\widetilde{S}^{(m)}\rightarrow P(\widetilde{S}%
^{(n)})$. If it has initial states, the function $\phi_{0}:\widetilde{S}%
^{(m)}\rightarrow P(\mathbf{B}^{n})$ that is defined by%
\[
\forall u\in\widetilde{S}^{(m)},\phi_{0}(u)=\{x(-\infty+0)|x\in f(u)\}
\]
is called the initial state function of $f$ and the set
\[
\Theta_{0}=\underset{u\in\widetilde{S}^{(m)}}{\bigcup}\phi_{0}(u)
\]
is called the set of the initial states of $f$.
\end{definition}

\begin{definition}
Let the pseudo-system $f$. If it has final states, the function $\phi
_{f}:\widetilde{S}^{(m)}\rightarrow P(\mathbf{B}^{n})$ that is given by%
\[
\forall u\in\widetilde{S}^{(m)},\phi_{f}(u)=\{x(\infty-0)|x\in f(u)\}
\]
is called the final state function of $f$ and the set
\[
\Theta_{f}=\underset{u\in\widetilde{S}^{(m)}}{\bigcup}\phi_{f}(u)
\]
is called the set of the final states of $f$.
\end{definition}

\begin{example}
The constant function $\widetilde{S}^{(m)}\rightarrow P(\widetilde{S}^{(n)})$
equal with $\{\mu\}$ is a pseudo-system with a constant initial state $\mu$
and fix initial time and it is also absolutely constantly stable with fix
final time. $\phi_{0},\Theta_{0},\phi_{f},\Theta_{f}$ are all defined and
equal with $\{\mu\}$.
\end{example}

\begin{theorem}
Let $f$ a pseudo-system with initial states.

a) If its initial states are race-free, then $\forall u\in\widetilde{S}%
^{(m)},\phi_{0}(u)$ has at most one element.

b) If $f$ has a constant initial state $\mu$, then $\phi_{0}(u)=\{\mu\}$ is
true for any admissible $u;$ for $f=\emptyset$ we have $\Theta_{0}=\emptyset$
and for $f\neq\emptyset$ we have $\Theta_{0}=\{\mu\}$.
\end{theorem}

\begin{proof}
a) We suppose that $f$ has race-free initial states and let $u\in\widetilde
{S}^{(m)}.$ If $f(u)=\emptyset$, then $\phi_{0}(u)=\emptyset$ and if
$f(u)\neq\emptyset$, then a unique $\mu\in\mathbf{B}^{n}$ exists, depending on
$u$ so that $\phi_{0}(u)=\{\mu\}$.

b) We suppose that $f$ has a constant initial state $\mu$. If $f$ is null then
$\forall u\in\widetilde{S}^{(m)},\phi_{0}(u)=\emptyset$ and $\Theta
_{0}=\emptyset$, otherwise for all admissible $u$ we have $\phi_{0}%
(u)=\{\mu\}$, the constant function thus $\Theta_{0}=\{\mu\}$.
\end{proof}

\begin{theorem}
We consider the pseudo-system $f$ with final states.

a) If its final states are race-free, then $\forall u\in\widetilde{S}%
^{(m)},\phi_{f}(u)$ has at most one element.

b) If $f$ has a constant final state $\mu$, then $\phi_{f}(u)=\{\mu\}$ is true
for any admissible $u;$ if admissible inputs do not exist then $\Theta
_{f}=\emptyset$ and if admissible inputs exist then $\Theta_{f}=\{\mu\}$.
\end{theorem}

\section{Pseudo-subsystems}

\begin{definition}
The pseudo-systems $f,g:\widetilde{S}^{(m)}\rightarrow P(\widetilde{S}^{(n)})$
are given. If%
\[
\forall u\in\widetilde{S}^{(m)},f(u)\subset g(u)
\]
then $f$ is called a pseudo-subsystem of $g$ and the usual notation is
$f\subset g$.
\end{definition}

\begin{remark}
Intuitively, the fact that $f$ is a pseudo-subsystem of $g$ shows that the
modeling of a circuit is made more precisely by $f$ than by $g$, by
considering a smaller set of admissible inputs, perhaps. $\subset$ is a
relation of partial order between $\widetilde{S}^{(m)}\rightarrow
P(\widetilde{S}^{(n)})$ pseudo-systems, where the first element is the null
pseudo-system and the last element is the total pseudo-system, see Remark 3.2.
\end{remark}

\begin{theorem}
Let the pseudo-system $g$ and $f\subset g$ an arbitrary pseudo-subsystem. If
$g$ has initial states (race-free initial states, constant initial state),
then $f$ has initial states (race-free initial states, constant initial state).
\end{theorem}

\begin{proof}
If one of the previous properties is true for the states in $g(u)$, then it is
true for the states in the subset $f(u)\subset g(u)$ also, $u\in\widetilde
{S}^{(m)}$.
\end{proof}

\begin{theorem}
Let $f\subset g$. If $g$ has final states (race-free final states, constant
final state), then $f$ has final states (race-free final states, constant
final state).
\end{theorem}

\begin{theorem}
The pseudo-systems $f\subset g$ are given. If $g$ has unbounded initial time
(bounded initial time, universal initial time), then $f$ has unbounded initial
time (bounded initial time, universal initial time).
\end{theorem}

\begin{proof}
Like previously, if one of the above properties is true for the states in
$g(u),$ then it is true for the states in $f(u)\subset g(u),u\in\widetilde
{S}^{(m)}$.
\end{proof}

\begin{theorem}
Let $f$ be a pseudo-subsystem of $g.$ If $g$ has unbounded final time (bounded
final time, universal final time), then $f$ has unbounded final time (bounded
final time, universal final time).
\end{theorem}

\begin{theorem}
If $g$ has initial states and $f\subset g$, then we note with $\gamma
_{0}:\widetilde{S}^{(m)}\rightarrow P(\mathbf{B}^{n})$ the initial state
function of $g$ and with $\Gamma_{0}\subset\mathbf{B}^{n}$ the set of the
initial states of $g$. We have $\forall u\in\widetilde{S}^{(m)},\phi
_{0}(u)\subset\gamma_{0}(u)$ and $\Theta_{0}\subset\Gamma_{0}$.
\end{theorem}

\begin{proof}
$f$ has initial states from Theorem 7.3, thus $\varphi_{0}$ and $\Theta_{0}$
exist. Moreover, as $\forall u\in\widetilde{S}^{(m)},f(u)\subset g(u)$, the
initial values of the states in $f(u)$ are contained between the initial
values of the states in $g(u)$, $\phi_{0}(u)\subset\gamma_{0}(u)$ making
$\Theta_{0}\subset\Gamma_{0}$ true too.
\end{proof}

\begin{theorem}
If $g$ has final states and $f\subset g$, we note with $\gamma_{f}%
:\widetilde{S}^{(m)}\rightarrow P(\mathbf{B}^{n})$ the final state function of
$g$ and with $\Gamma_{f}\subset\mathbf{B}^{n}$ the set of the final states of
$g$. We have $\forall u\in\widetilde{S}^{(m)},\phi_{f}(u)\subset\gamma_{f}(u)$
and $\Theta_{f}\subset\Gamma_{f}$.
\end{theorem}

\section{Dual pseudo-systems}

\begin{notation}
For any $\lambda\in\mathbf{B}^{m},u\in\widetilde{S}^{(m)}$ we note with
$\overline{\lambda}\in\mathbf{B}^{m},\overline{u}\in\widetilde{S}^{(m)}$ the
complements of $\lambda,u$ made coordinatewise:%
\[
\overline{\lambda}=(\overline{\lambda_{1}},...,\overline{\lambda_{m}})
\]%
\[
\overline{u}(t)=(\overline{u_{1}(t)},...,\overline{u_{m}(t)})
\]
\end{notation}

\begin{definition}
Let the pseudo-system $f:\widetilde{S}^{(m)}\rightarrow P(\widetilde{S}%
^{(n)})$. The pseudo-system $f^{\ast}:\widetilde{S}^{(m)}\rightarrow
P(\widetilde{S}^{(n)})$ that is defined by%
\[
\forall u\in\widetilde{S}^{(m)},f^{\ast}(u)=\{\overline{x}|x\in f(\overline
{u})\}
\]
is called the dual pseudo-system of $f$.
\end{definition}

\begin{remark}
We add to the types of duality that were previously presented the duality
between $0,1\in\mathbf{B}$ that gives Definition 8.2. The dual pseudo-system
$f^{\ast}$ has many properties that can be inferred from those of $f$.
\end{remark}

\begin{theorem}
$(f^{\ast})^{\ast}=f$.
\end{theorem}

\begin{theorem}
The next statements are equivalent for the pseudo-system $f$:

a) $f$ has initial states (race-free initial states, constant initial state)

b) $f^{\ast}$ has initial states (race-free initial states, constant initial state).
\end{theorem}

\begin{proof}
We show that $f$ has race-free initial states $\Longleftrightarrow$ $f^{\ast}$
has race-free initial states:%
\[
\forall u\in\widetilde{S}^{(m)},\exists\mu\in\mathbf{B}^{n},\forall x\in
f(u),\exists t_{0}\in\mathbf{R},\forall t<t_{0},x(t)=\mu\Longleftrightarrow
\]%
\[
\forall u\in\widetilde{S}^{(m)},\exists\mu\in\mathbf{B}^{n},\forall
\overline{x}\in f^{\ast}(\overline{u}),\exists t_{0}\in\mathbf{R},\forall
t<t_{0},x(t)=\mu\Longleftrightarrow
\]%
\[
\forall\overline{u}\in\widetilde{S}^{(m)},\exists\overline{\mu}\in
\mathbf{B}^{n},\forall\overline{x}\in f^{\ast}(\overline{u}),\exists t_{0}%
\in\mathbf{R},\forall t<t_{0},\overline{x}(t)=\overline{\mu}%
\Longleftrightarrow
\]%
\[
\forall u\in\widetilde{S}^{(m)},\exists\mu\in\mathbf{B}^{n},\forall x\in
f^{\ast}(u),\exists t_{0}\in\mathbf{R},\forall t<t_{0},x(t)=\mu
\]
\end{proof}

\begin{theorem}
For the pseudo-system $f$, the next statements are equivalent:

a) $f$ has final states (race-free final states, constant final state)

b) $f^{\ast}$ has final states (race-free final states, constant final state).
\end{theorem}

\begin{theorem}
The next properties are equivalent for $f:$

a) $f$ has unbounded initial time (bounded initial time, fix initial time)

b) $f^{\ast}$ has unbounded initial time (bounded initial time, fix initial time).
\end{theorem}

\begin{theorem}
Let the pseudo-system $f$. The next properties are equivalent:

a) $f$ has unbounded final time (bounded final time, fix final time)

b) $f^{\ast}$ has unbounded final time (bounded final time, fix final time).
\end{theorem}

\begin{theorem}
If $f$ has initial states, we note with $\phi_{0}^{\ast}:\widetilde{S}%
^{(m)}\rightarrow P(\mathbf{B}^{n})$ the initial state function of $f^{\ast}$
and with $\Theta_{0}^{\ast}$ the set of the initial states of $f^{\ast}$. We
have%
\[
\forall u\in\widetilde{S}^{(m)},\phi_{0}^{\ast}(u)=\{\overline{\mu}|\mu\in
\phi_{0}(\overline{u})\}
\]%
\[
\Theta_{0}^{\ast}=\{\overline{\mu}|\mu\in\Theta_{0}^{{}}\}
\]
\end{theorem}

\begin{proof}
If $f$ has initial states, then $f^{\ast}$ has initial states (Theorem 8.5)
thus $\phi_{0}^{\ast}$ and $\Theta_{0}^{\ast}$ exist. The statements of the
theorem are obtained from the fact that%
\[
\forall u\in\widetilde{S}^{(m)},\phi_{0}^{\ast}(u)=\{x(-\infty+0)|x\in
f^{\ast}(u)\}=
\]%
\[
=\{\overline{x(-\infty+0)}|\overline{x}\in f^{\ast}(u)\}=\{\overline
{x(-\infty+0)}|x\in f(\overline{u})\}=\{\overline{\mu}|\mu\in\phi
_{0}(\overline{u})\}
\]
\end{proof}

\begin{theorem}
If $f$ has final states, we note with $\phi_{f}^{\ast}:\widetilde{S}%
^{(m)}\rightarrow P(\mathbf{B}^{n})$ the final state function of $f^{\ast}$
and with $\Theta_{f}^{\ast}$ the set of the final states of $f^{\ast}$. We
have%
\[
\forall u\in\widetilde{S}^{(m)},\phi_{f}^{\ast}(u)=\{\overline{\mu}|\mu\in
\phi_{f}(\overline{u})\}
\]%
\[
\Theta_{f}^{\ast}=\{\overline{\mu}|\mu\in\Theta_{f}^{{}}\}
\]
\end{theorem}

\begin{theorem}
For the pseudo-systems $f,g:\widetilde{S}^{(m)}\rightarrow P(\widetilde
{S}^{(n)})$ we have $f\subset g\Longleftrightarrow f^{\ast}\subset g^{\ast}$
\end{theorem}

\begin{proof}
We get the next sequence of equivalencies:%
\[
\forall u\in\widetilde{S}^{(m)},f(u)\subset g(u)\Longleftrightarrow\forall
u\in\widetilde{S}^{(m)},\{x|x\in f(u)\}\subset\{x|x\in
g(u)\}\Longleftrightarrow
\]%
\[
\Longleftrightarrow\forall u\in\widetilde{S}^{(m)},\{\overline{x}|x\in
f(u)\}\subset\{\overline{x}|x\in g(u)\}\Longleftrightarrow\forall
u\in\widetilde{S}^{(m)},f^{\ast}(\overline{u})\subset g^{\ast}(\overline
{u})\Longleftrightarrow
\]%
\[
\Longleftrightarrow\forall\overline{u}\in\widetilde{S}^{(m)},f^{\ast
}(u)\subset g^{\ast}(u)\Longleftrightarrow\forall u\in\widetilde{S}%
^{(m)},f^{\ast}(u)\subset g^{\ast}(u)
\]
\end{proof}

\section{Inverse pseudo-systems}

\begin{definition}
Let $f:\widetilde{S}^{(m)}\rightarrow P(\widetilde{S}^{(n)})$. The
pseudo-system $f^{-1}:\widetilde{S}^{(n)}\rightarrow P(\widetilde{S}^{(m)})$,
called the inverse of $f$, is defined by%
\[
\forall x\in\widetilde{S}^{(n)},f^{-1}(x)=\{u|u\in\widetilde{S}^{(m)},x\in
f(u)\}
\]
\end{definition}

\begin{remark}
The idea of construction of $f^{-1}$ is that of inverting the cause-effect
relation: it associates to each possible effect $x$ these admissible inputs
$u$ that could have caused it. We observe that $u\in f^{-1}%
(x)\Longleftrightarrow x\in f(u)$.
\end{remark}

\begin{example}
The inverse of the null pseudo-system $f$ is the null pseudo-system and the
inverse of the total pseudo-system is the total pseudo-system.
\end{example}

\begin{theorem}
For the pseudo-system $f$ we have $(f^{-1})^{-1}=f$.
\end{theorem}

\begin{proof}
For any $\forall u\in\widetilde{S}^{(m)}$, we can write that
\[
(f^{-1})^{-1}(u)=\{x|u\in f^{-1}(x)\}=\{x|x\in f(u)\}=f(u)
\]
\end{proof}

\begin{theorem}
If $f^{-1}$ has initial states, then the admissible inputs of $f$ are signals.
\end{theorem}

\begin{proof}
We suppose the contrary, i.e. some admissible input $u^{0}$ of $f$ exists that
is not a signal:%
\[
u^{0}\in\widetilde{S}^{(m)}\quad and\quad f(u^{0})\neq\emptyset\quad
and\quad\daleth(\exists\lambda\in\mathbf{B}^{m},\exists t_{0}\in
\mathbf{R},\forall t<t_{0},u(t)=\lambda)
\]
We take some $x^{0}\in f(u^{0})$, meaning that $u^{0}\in f^{-1}(x^{0})$. In
the statement relative to the initial states of $f^{-1}$:%
\[
\forall x\in\widetilde{S}^{(n)},\forall u\in f^{-1}(x),\exists\lambda
\in\mathbf{B}^{m},\exists t_{0}\in\mathbf{R},\forall t<t_{0},u(t)=\lambda
\]
we have for $x=x^{0}$ and $u=u^{0}$:%
\[
x^{0}\in\widetilde{S}^{(n)}\Longrightarrow(u^{0}\in f^{-1}(x^{0}%
)\Longrightarrow\exists\lambda\in\mathbf{B}^{m},\exists t_{0}\in
\mathbf{R},\forall t<t_{0},u(t)=\lambda)
\]
The two prerequisites are true and the conclusion is false, contradiction.
\end{proof}

\begin{theorem}
We suppose that $f^{-1}$ has initial states and we note with $\phi_{0}%
^{-1}:\widetilde{S}^{(n)}\rightarrow P(\mathbf{B}^{m}),$ $\Theta_{0}^{-1}$ its
initial state function and respectively its set of initial states. We have%
\[
\forall x\in\widetilde{S}^{(n)},\phi_{0}^{-1}(x)=\{u(-\infty+0)|u\in
\widetilde{S}^{(m)},x\in f(u)\}
\]%
\[
\Theta_{0}^{-1}=\{u(-\infty+0)|u\in\widetilde{S}^{(m)},f(u)\neq\emptyset\}
\]
\end{theorem}

\begin{theorem}
We suppose that $f^{-1}$ has final states and we note with $\phi_{f}%
^{-1}:\widetilde{S}^{(n)}\rightarrow P(\mathbf{B}^{m}),$ $\Theta_{f}^{-1}$ its
final state function and respectively its set of final states. We have%
\[
\forall x\in\widetilde{S}^{(n)},\phi_{f}^{-1}(x)=\{u(\infty-0)|u\in
\widetilde{S}^{(m)},x\in f(u)\}
\]%
\[
\Theta_{f}^{-1}=\{u(\infty-0)|u\in\widetilde{S}^{(m)},f(u)\neq\emptyset\}
\]
\end{theorem}

\begin{theorem}
If $f\subset g$, then $f^{-1}\subset g^{-1}$ and $(f^{\ast})^{-1}%
\subset(g^{\ast})^{-1}$ take place.
\end{theorem}

\begin{proof}
$\forall u\in\widetilde{S}^{(m)},f(u)\subset g(u)$ implies%
\[
\forall u\in\widetilde{S}^{(m)},\forall x\in\widetilde{S}^{(n)},x\in
f(u)\Longrightarrow x\in g(u)
\]%
\[
\forall u\in\widetilde{S}^{(m)},\forall x\in\widetilde{S}^{(n)},u\in
f^{-1}(x)\Longrightarrow u\in g^{-1}(x)
\]%
\[
\forall x\in\widetilde{S}^{(n)},\forall u\in\widetilde{S}^{(m)},u\in
f^{-1}(x)\Longrightarrow u\in g^{-1}(x)
\]%
\[
\forall x\in\widetilde{S}^{(n)},f^{-1}(x)\subset g^{-1}(x)
\]
On the other hand $f\subset g$ implies $f^{\ast}\subset g^{\ast}$ (see Theorem
8.11) and from the previous item we get $(f^{\ast})^{-1}\subset(g^{\ast}%
)^{-1}$.
\end{proof}

\begin{theorem}
$(f^{-1})^{\ast}=(f^{\ast})^{-1}$.
\end{theorem}

\begin{proof}
We get \ for all $x\in\widetilde{S}^{(n)}$ that%
\[
(f^{-1})^{\ast}(x)=\{\overline{u}|u\in f^{-1}(\overline{x})\}=\{\overline
{u}|\overline{x}\in f(u)\}=\{\overline{u}|x\in f^{\ast}(\overline{u})\}=
\]%
\[
=\{u|x\in f^{\ast}(u)\}=\{u|u\in(f^{\ast})^{-1}(x)\}=(f^{\ast})^{-1}(x)
\]
\end{proof}

\section{Direct product}

\begin{definition}
We consider the pseudo-systems $f:\widetilde{S}^{(m)}\rightarrow
P(\widetilde{S}^{(n)}),$ $f^{\prime}:\widetilde{S}^{(m^{\prime})}\rightarrow
P(\widetilde{S}^{(n^{\prime})})$. The direct product of $f$ and $f^{\prime}$
is by definition the pseudo-system $f\times f^{\prime}:\widetilde
{S}^{(m+m^{\prime})}\rightarrow P(\widetilde{S}^{(n+n^{\prime})})$ that is
defined in the next manner:%
\[
\forall(u,u^{\prime})\in\widetilde{S}^{(m+m^{\prime})},(f\times f^{\prime
})(u,u^{\prime})=\{(x,x^{\prime})|(x,x^{\prime})\in\widetilde{S}%
^{(n+n^{\prime})},x\in f(u),x^{\prime}\in f^{\prime}(u^{\prime})\}
\]
where $u$ is the projection of the variable from $\widetilde{S}^{(m+m^{\prime
})}$ on the first $m$ coordinates and $u^{\prime}$ is the projection of the
variable from $\widetilde{S}^{(m+m^{\prime})}$ on the last $m^{\prime}$
coordinates. Similarly, $x$ is the projection of the variable from
$\widetilde{S}^{(n+n^{\prime})}$ on the first $n$ coordinates and $x^{\prime}$
is the projection of the same variable on the last $n^{\prime}$ coordinates.
\end{definition}

\begin{remark}
$f\times f^{\prime}$ is the pseudo-system representing $f$ and $f^{\prime}$
acting independently on each other. Some sort of problem arises here, from the
fact that 'independently on each other' refers to the function $f\times
f^{\prime}:\widetilde{S}^{(m)}\times\widetilde{S}^{(m^{\prime})}\rightarrow
P(\widetilde{S}^{(n)})\times P(\widetilde{S}^{(n^{\prime})})$ and we were
forced to make the identifications between $\widetilde{S}^{(m)}\times
\widetilde{S}^{(m^{\prime})}$ and $\widetilde{S}^{(m+m^{\prime})}$ and
respectively between $P(\widetilde{S}^{(n)})\times P(\widetilde{S}%
^{(n^{\prime})})$ and $P(\widetilde{S}^{(n+n^{\prime})})$, in order that
$f\times f^{\prime}$ be a pseudo-system. This means exactly one time axis
(like in $\widetilde{S}^{(m+m^{\prime})}$and $P(\widetilde{S}^{(n+n^{\prime}%
)})$) instead of two (like in $\widetilde{S}^{(m)}\times\widetilde
{S}^{(m^{\prime})}$ and $P(\widetilde{S}^{(n)})\times P(\widetilde
{S}^{(n^{\prime})})$). But in this moment $f$ and $f^{\prime}$ do not quite
act 'independently on each other'. Things look like claiming 'time is
universal, the same for everybody'.

On the other hand, we can write%
\[
\forall(u,u^{\prime})\in\widetilde{S}^{(m+m^{\prime})},(f\times f^{\prime
})(u,u^{\prime})=f(u)\times f^{\prime}(u^{\prime})
\]
if we accept that the elements of $f(u)\times f^{\prime}(u^{\prime})$ belong
to $P(\widetilde{S}^{(n+n^{\prime})})$ (not to $P(\widetilde{S}^{(n)})\times
P(\widetilde{S}^{(n^{\prime})})$).
\end{remark}

\begin{theorem}
The pseudo-systems $f$ and $f^{\prime}$ have initial states (race-free initial
states, constant initial state) if and only if $f\times f^{\prime}$ has
initial states (race-free initial states, constant initial state).
\end{theorem}

\begin{proof}
For example the conjunction of the statements%
\[
\exists\mu\in\mathbf{B}^{n},\forall u\in\widetilde{S}^{(m)},\forall x\in
f(u),\exists t_{0}\in\mathbf{R},\forall t<t_{0},x(t)=\mu
\]%
\[
\exists\mu^{\prime}\in\mathbf{B}^{n^{\prime}},\forall u^{\prime}\in
\widetilde{S}^{(m^{\prime})},\forall x^{\prime}\in f^{\prime}(u^{\prime
}),\exists t_{0}^{^{\prime}}\in\mathbf{R},\forall t<t_{0}^{^{\prime}%
},x^{\prime}(t)=\mu^{\prime}%
\]
is equivalent with%
\[
\exists(\mu,\mu^{\prime})\in\mathbf{B}^{n+n^{\prime}},\forall(u,u^{\prime}%
)\in\widetilde{S}^{(m+m^{\prime})},\forall(x,x^{\prime})\in(f\times f^{\prime
})(u,u^{\prime}),
\]%
\[
\exists t_{0}^{"}\in\mathbf{R},\forall t<t_{0}^{"},(x(t),x^{\prime}%
(t))=(\mu,\mu^{\prime})
\]
where we can take $t_{0}^{"}=\min(t_{0},t_{0}^{\prime})$ each time.
\end{proof}

\begin{theorem}
$f$ and $f^{\prime}$ have final states (race-free final states, constant final
state) if and only if $f\times f^{\prime}$ has final states (race-free final
states, constant final state).
\end{theorem}

\begin{theorem}
Let the pseudo-systems $f,f^{\prime}$. The next statements are equivalent:

a) $f$ and $f^{\prime}$ have unbounded initial time (bounded initial time, fix
initial time)

b) $f\times f^{\prime}$ has unbounded initial time (bounded initial time, fix
initial time).
\end{theorem}

\begin{theorem}
The pseudo-systems $f$ and $f^{\prime}$ have unbounded final time (bounded
final time, fix final time) if and only if $f\times f^{\prime}$ has unbounded
final time (bounded final time, fix final time).
\end{theorem}

\begin{theorem}
Let the pseudo-systems $f$ and $f^{\prime}$ defined like before. If they have
initial states, we note with $\phi_{0},\phi_{0}^{\prime}$ their initial state
functions and with $(\phi\times\phi^{\prime})_{0}:\widetilde{S}^{(m+m^{\prime
})}\rightarrow P(\mathbf{B}^{n+n^{\prime}})$ the initial state function of
$f\times f^{\prime}$. We also note with $\Theta_{0}^{{}},\Theta_{0}^{^{\prime
}}$ the sets of the initial states of $f$ and $f^{\prime}$ and let
$(\Theta\times\Theta^{\prime})_{0}$ the set of the initial states of $f\times
f^{\prime}$. We have%
\[
\forall(u,u^{\prime})\in\widetilde{S}^{(m+m^{\prime})},(\phi\times\phi
^{\prime})_{0}(u,u^{\prime})=\phi_{0}(u)\times\phi_{0}^{\prime}(u)
\]%
\[
(\Theta\times\Theta^{\prime})_{0}=\Theta_{0}\times\Theta_{0}^{^{\prime}}%
\]
In the previous equations we have identified $P(\mathbf{B}^{n})\times
P(\mathbf{B}^{n^{\prime}})$ with $P(\mathbf{B}^{n+n^{\prime}})$.
\end{theorem}

\begin{proof}
If $f,f^{\prime}$ have initial states, then $f\times f^{\prime}$ has initial
states from Theorem 10.3 thus $(\phi\times\phi^{\prime})_{0}$ and
$(\Theta\times\Theta^{\prime})_{0}$ exist. We obtain%
\[
\forall(u,u^{\prime})\in\widetilde{S}^{(m+m^{\prime})},(\phi\times\phi
^{\prime})_{0}(u,u^{\prime})=\{(x(-\infty+0),x^{\prime}(-\infty
+0))|(x,x^{\prime})\in(f\times f^{\prime})(u,u^{\prime})\}=
\]%
\[
=\{(x(-\infty+0),x^{\prime}(-\infty+0))|x\in f(u),x^{\prime}\in f^{\prime
}(u^{\prime})\}=
\]%
\[
=\{x(-\infty+0)|x\in f(u)\}\times\{x^{\prime}(-\infty+0)|x^{\prime}\in
f^{\prime}(u^{\prime})\}=\phi_{0}(u)\times\phi_{0}^{\prime}(u^{\prime})
\]%
\[
(\Theta\times\Theta^{\prime})_{0}=\underset{(u,u^{\prime})\in\widetilde
{S}^{(m+m^{\prime})}}{\bigcup}(\phi\times\phi^{\prime})_{0}(u,u^{\prime
})=\underset{(u,u^{\prime})\in\widetilde{S}^{(m+m^{\prime})}}{\bigcup}\phi
_{0}(u)\times\phi_{0}^{\prime}(u^{\prime})=
\]%
\[
=\underset{u\in\widetilde{S}^{(m)}}{\bigcup}\phi_{0}(u)\times\underset
{u^{\prime}\in\widetilde{S}^{(m^{\prime})}}{\bigcup}\phi_{0}^{\prime
}(u^{\prime})=\Theta_{0}\times\Theta_{0}^{^{\prime}}%
\]
\end{proof}

\begin{theorem}
If $f$,$f^{\prime}$ have final states, we note with $\phi_{f},\phi_{f}%
^{\prime}$ their final state functions and with $(\phi\times\phi^{\prime}%
)_{f}:\widetilde{S}^{(m+m^{\prime})}\rightarrow P(\mathbf{B}^{n+n^{\prime}})$
the final state function of $f\times f^{\prime}$. We also note with
$\Theta_{f}^{{}},\Theta_{f}^{^{\prime}}$ the sets of the final states of $f$
and $f^{\prime}$ and with $(\Theta\times\Theta^{\prime})_{f}$ the set of the
final states of $f\times f^{\prime}$. We have%
\[
\forall(u,u^{\prime})\in\widetilde{S}^{(m+m^{\prime})},(\phi\times\phi
^{\prime})_{f}(u,u^{\prime})=\phi_{f}(u)\times\phi_{f}^{\prime}(u^{\prime})
\]%
\[
(\Theta\times\Theta^{\prime})_{f}=\Theta_{f}\times\Theta_{f}^{^{\prime}}%
\]
and the same identification between $P(\mathbf{B}^{n})\times P(\mathbf{B}%
^{n^{\prime}})$ and $P(\mathbf{B}^{n+n^{\prime}})$ like before has been made.
\end{theorem}

\begin{theorem}
Let the pseudo-systems $f,g:\widetilde{S}^{(m)}\rightarrow P(\widetilde
{S}^{(n)})$, $f^{\prime},g^{\prime}:\widetilde{S}^{(m^{\prime})}\rightarrow
P(\widetilde{S}^{(n^{\prime})})$. We have that $f\subset g$ and $f^{\prime
}\subset g^{\prime}$ if and only if $f\times f^{\prime}\subset g\times
g^{\prime}$.
\end{theorem}

\begin{theorem}
For any pseudo-systems $f,f^{\prime}$ we have $(f\times f^{\prime})^{\ast
}=f^{\ast}\times f^{\prime\ast}$.
\end{theorem}

\begin{proof}
For any $(u,u^{\prime})\in\widetilde{S}^{(m+m^{\prime})}$ we can write%
\[
(f\times f^{\prime})^{\ast}(u,u^{\prime})=\{(\overline{x},\overline{x^{\prime
}})|(x,x^{\prime})\in(f\times f^{\prime})(\overline{u},\overline{u^{\prime}%
})\}=\{(\overline{x},\overline{x^{\prime}})|x\in f(\overline{u}),x^{\prime}\in
f^{\prime}(\overline{u^{\prime}})\}=
\]%
\[
=\{(\overline{x},\overline{x^{\prime}})|\overline{x}\in f^{\ast}%
(u),\overline{x^{\prime}}\in f^{\prime\ast}(u^{\prime})\}=\{(x,x^{\prime
})|x\in f^{\ast}(u),x^{\prime}\in f^{\prime\ast}(u^{\prime})\}=
\]%
\[
=\{(x,x^{\prime})|(x,x^{\prime})\in(f^{\ast}\times f^{\prime\ast}%
)(u,u^{\prime})\}=(f^{\ast}\times f^{\prime\ast})(u,u^{\prime})
\]
\end{proof}

\begin{theorem}
Let $f$ and $f^{\prime}$. We have that $(f\times f^{\prime})^{-1}=f^{-1}\times
f^{\prime-1}$.
\end{theorem}

\section{Parallel connection}

\begin{definition}
The pseudo-systems $f:\widetilde{S}^{(m)}\rightarrow P(\widetilde{S}^{(n)})$
and $f^{\prime}:\widetilde{S}^{(m)}\rightarrow P(\widetilde{S}^{(n^{\prime}%
)})$ are considered. The pseudo-system $(f,f^{\prime}):\widetilde{S}%
^{(m)}\rightarrow P(S^{(n+n^{\prime})})$ that is defined in the next manner%
\[
\forall u\in\widetilde{S}^{(m)},(f,f^{\prime})(u)=\{(x,x^{\prime
})|(x,x^{\prime})\in\widetilde{S}^{(n+n^{\prime})},x\in f(u),x^{\prime}\in
f^{\prime}(u)\}
\]
is called the parallel connection of the systems $f$ and $f^{\prime}$.
\end{definition}

\begin{remark}
The study of the parallel connection of the pseudo-systems is made in quite
similar terms with the study of the direct product of pseudo-systems from the
previous section.

The relation between the direct product and the parallel connection is
expressed by the commutativity of the next diagram%
\[%
\begin{array}
[c]{ccc}%
\begin{array}
[c]{c}%
\\
\widetilde{S}^{(m)}%
\end{array}
& \underrightarrow{\quad(f,f^{\prime})\quad} &
\begin{array}
[c]{c}%
\\
P(\widetilde{S}^{(n+n^{\prime})})
\end{array}
\\
\Delta\downarrow\quad &  & |\,|\\
\widetilde{S}^{(2m)} &
\begin{array}
[c]{c}%
\\
\overrightarrow{\quad f\times f^{\prime}\quad}%
\end{array}
& P(\widetilde{S}^{(n+n^{\prime})})
\end{array}
\]
where we have noted with $\Delta$ the diagonal function%
\[
\forall u\in\widetilde{S}^{(m)},\Delta(u)=(u,u)
\]
\end{remark}

\section{Serial connection}

\begin{definition}
Let the pseudo-systems $f:\widetilde{S}^{(m)}\rightarrow P(\widetilde{S}%
^{(n)})$ and $h:\widetilde{S}^{(n)}\rightarrow P(\widetilde{S}^{(p)})$. The
pseudo-system $h\circ f:\widetilde{S}^{(m)}\rightarrow P(\widetilde{S}^{(p)})$
that is defined in the next way%
\[
\forall u\in\widetilde{S}^{(m)},(h\circ f)(u)=\{y|\exists x\in f(u),y\in
h(x)\}
\]
is called the serial connection of the pseudo-systems $h$ and $f$.
\end{definition}

\begin{theorem}
Let the pseudo-systems $f,h$. If $h$ has initial states (constant initial
state), then $h\circ f$ has initial states (constant initial state).
\end{theorem}

\begin{proof}
For example from%
\[
\forall u\in\widetilde{S}^{(m)},\forall y\in(h\circ f)(u),\exists x\in
f(u),y\in h(x)
\]%
\[
\exists\nu\in\mathbf{B}^{p},\forall x\in\widetilde{S}^{(n)},\forall y\in
h(x),\exists t_{0}\in\mathbf{R},\forall t<t_{0},y(t)=\nu
\]
we infer%
\[
\exists\nu\in\mathbf{B}^{p},\forall u\in\widetilde{S}^{(m)},\forall
y\in(h\circ f)(u),\exists t_{0}\in\mathbf{R},\forall t<t_{0},y(t)=\nu
\]
\end{proof}

\begin{theorem}
If $h$ has final states (constant final state), then $h\circ f$ has final
states (constant final state).
\end{theorem}

\begin{theorem}
Let the systems $f$ and $h$. If $h$ has unbounded initial time (fix initial
time), then $h\circ f$ has unbounded initial time (fix initial time).
\end{theorem}

\begin{proof}
For example from%
\[
\forall u\in\widetilde{S}^{(m)},\forall y\in(h\circ f)(u)\cap S^{(p)},\exists
x\in f(u),y\in h(x)\cap S^{(p)}%
\]%
\[
\exists t_{0}\in\mathbf{R},\forall x\in\widetilde{S}^{(n)},\forall y\in
h(x)\cap S^{(p)},\exists\nu\in\mathbf{B}^{p},\forall t<t_{0},y(t)=\nu
\]
we get%
\[
\exists t_{0}\in\mathbf{R},\forall u\in\widetilde{S}^{(m)},\forall y\in(h\circ
f)(u)\cap S^{(p)},\exists\nu\in\mathbf{B}^{p},\forall t<t_{0},y(t)=\nu
\]
\end{proof}

\begin{theorem}
If $h$ has unbounded final time (fix final time), then $h\circ f$ has
unbounded final time (fix final time).
\end{theorem}

\begin{theorem}
We consider the pseudo-systems $f$ and $h$. If $h$ has initial states, we note
with $\varphi_{0},\delta_{0}$ on one hand and $\Delta_{0}$ on the other hand
the initial state functions of $h,h\circ f$, respectively the set of initial
states of $h\circ f$. The next formulas are true:%
\[
\forall u\in\widetilde{S}^{(m)},\delta_{0}(u)=\underset{x\in f(u)}{\bigcup
}\varphi_{0}(x)
\]%
\[
\Delta_{0}=\underset{u\in\widetilde{S}^{(m)}}{\bigcup}\text{ }\underset{x\in
f(u)}{\bigcup}\varphi_{0}(x)
\]
\end{theorem}

\begin{theorem}
Let the pseudo-systems $f$ and $h$. We suppose that $h$ has final states and
we use the notations $\varphi_{f},\delta_{f}$ on one hand and $\Delta_{f}$ on
the other hand for the final state functions of $h,h\circ f$, respectively for
the set of final states of $h\circ f$. The next formulas are true:%
\[
\forall u\in\widetilde{S}^{(m)},\delta_{f}(u)=\underset{x\in f(u)}{\bigcup
}\varphi_{f}(x)
\]%
\[
\Delta_{f}=\underset{u\in\widetilde{S}^{(m)}}{\bigcup}\text{ }\underset{x\in
f(u)}{\bigcup}\varphi_{f}(x)
\]
\end{theorem}

\begin{proof}
From the fact that $h$ has final states we infer, see Theorem 12.3, that
$h\circ f$ has final states so that $\delta_{f},\Delta_{f}$ exist. We have:%
\[
\forall u\in\widetilde{S}^{(m)},\delta_{f}(u)=\{y(\infty-0)|y\in(h\circ
f)(u)\}=\{y(\infty-0)|\exists x,x\in f(u)\quad and\quad y\in h(x)\}=
\]%
\[
=\underset{x\in f(u)}{\bigcup}\{y(\infty-0)|y\in h(x)\}=\underset{x\in
f(u)}{\bigcup}\varphi_{f}(x)
\]
\end{proof}

\begin{theorem}
Let's consider the pseudo-systems $f,g:\widetilde{S}^{(m)}\rightarrow
P(\widetilde{S}^{(n)})$ and $h,h_{1}:\widetilde{S}^{(n)}\rightarrow
P(\widetilde{S}^{(p)})$. We have:

a) $f\subset g\Longrightarrow h\circ f\subset h\circ g$

b) $h\subset h_{1}\Longrightarrow h\circ f\subset h_{1}\circ f$
\end{theorem}

\begin{proof}
Let $u\in\widetilde{S}^{(m)}$ arbitrary. Because $f(u)\subset g(u)$ we infer
that%
\[
(h\circ f)(u)=\{y|\exists x\in f(u),y\in h(x)\}\subset\{y|\exists x\in
g(u),y\in h(x)\}=(h\circ g)(u)
\]
\end{proof}

\begin{theorem}
For the pseudo-systems $f$ and $h$, we have $(h\circ f)^{\ast}=h^{\ast}\circ
f^{\ast}$.
\end{theorem}

\begin{proof}
For any $u\in\widetilde{S}^{(m)}$ we have%
\[
(h\circ f)^{\ast}(u)=\{\overline{y}|y\in(h\circ f)(\overline{u})\}=\{\overline
{y}|\exists x\in f(\overline{u}),y\in h(x)\}=
\]%
\[
=\{y|\exists\overline{x}\in f(\overline{u}),\overline{y}\in h(\overline
{x})\}=\{y|\exists x\in f^{\ast}(u),y\in h^{\ast}(x)\}=(h^{\ast}\circ f^{\ast
})(u)
\]
\end{proof}

\begin{theorem}
For any pseudo-system $f$ we have
\[
\forall u\in\widetilde{S}^{(m)},(f^{-1}\circ f)(u)=\{v|v\in\widetilde{S}%
^{(m)},f(u)\cap f(v)\neq\emptyset\}
\]%
\[
\forall x\in\widetilde{S}^{(n)},(f\circ f^{-1})(x)=\{z|z\in\widetilde{S}%
^{(n)},f^{-1}(x)\cap f^{-1}(z)\neq\emptyset\}
\]
\end{theorem}

\begin{proof}
We observe that%
\[
\forall u\in\widetilde{S}^{(m)},(f^{-1}\circ f)(u)=\{v|\exists x\in f(u),v\in
f^{-1}(x)\}=
\]%
\[
=\{v|\exists x,x\in f(u),x\in f(v)\}=\{v|f(u)\cap f(v)\neq\emptyset\}
\]
and similarly for the other statement.
\end{proof}

\begin{theorem}
Let $f$ and $h$. We have $(h\circ f)^{-1}=f^{-1}\circ h^{-1}$.
\end{theorem}

\begin{proof}
For any $y\in\widetilde{S}^{(p)}$ we have%
\[
(h\circ f)^{-1}(y)=\{u|y\in(h\circ f)(u)\}=\{u|\exists x,x\in f(u)\quad
and\quad y\in h(x)\}=
\]%
\[
=\{u|\exists x,x\in h^{-1}(y)\quad and\quad u\in f^{-1}(x)\}=(f^{-1}\circ
h^{-1})(y)
\]
\end{proof}

\begin{theorem}
We consider the pseudo-systems $f:\widetilde{S}^{(m)}\rightarrow
P(\widetilde{S}^{(n)})$, $f^{\prime}:\widetilde{S}^{(m^{\prime})}\rightarrow
P(\widetilde{S}^{(n^{\prime})})$, respectively $h:\widetilde{S}^{(n)}%
\rightarrow P(\widetilde{S}^{(p)})$, $h^{\prime}:\widetilde{S}^{(n^{\prime}%
)}\rightarrow P(\widetilde{S}^{(p^{\prime})})$. The next formula is true:%
\[
(h\times h^{\prime})\circ(f\times f^{\prime})=(h\circ f)\times(h^{\prime}\circ
f^{\prime})
\]
\end{theorem}

\begin{theorem}
If in the hypothesis of the previous theorem we have the special case
$m=m^{\prime}$, then we can write%
\[
(h\times h^{\prime})\circ(f,f^{\prime})=(h\circ f,h^{\prime}\circ f^{\prime})
\]
\end{theorem}

\begin{proof}
For any $u\in\widetilde{S}^{(m)}$ we have%
\[
((h\times h^{\prime})\circ(f,f^{\prime}))(u)=\{(y,y^{\prime})|\exists
(x,x^{\prime}),(x,x^{\prime})\in(f,f^{\prime})(u)\quad and\quad(y,y^{\prime
})\in(h\times h^{\prime})(x,x^{\prime})\}=
\]%
\[
=\{(y,y^{\prime})|\exists x,x\in f(u)\quad and\quad y\in h(x)\quad
and\quad\exists x^{\prime},x^{\prime}\in f^{\prime}(u)\quad and\quad
y^{\prime}\in h^{\prime}(x^{\prime})\}=
\]%
\[
=\{(y,y^{\prime})|y\in(h\circ f)(u)\quad and\quad y^{\prime}\in(h^{\prime
}\circ f^{\prime})(u)\}=(h\circ f,h^{\prime}\circ f^{\prime})(u)
\]
\end{proof}

\section{Complement}

\begin{definition}
Let $f:\widetilde{S}^{(m)}\rightarrow P(\widetilde{S}^{(n)})$. The
pseudo-system $Cf:\widetilde{S}^{(m)}\rightarrow P(\widetilde{S}^{(n)})$ that
is defined by%
\[
\forall u\in\widetilde{S}^{(m)},Cf(u)=\widetilde{S}^{(n)}\setminus f(u)
\]
is called the complement of $f$.
\end{definition}

\begin{remark}
Intuitively, if $x\in f(u)$ are these states that model a circuit then $x\in
Cf(u)$ are the states that do not model that circuit.
\end{remark}

\begin{theorem}
$CCf=f$
\end{theorem}

\begin{theorem}
If $f,g:\widetilde{S}^{(m)}\rightarrow P(\widetilde{S}^{(n)}),$ then $f\subset
g$ if and only if $Cg\subset Cf$
\end{theorem}

\begin{proof}
We have%
\[
f\subset g\Longleftrightarrow\forall u\in\widetilde{S}^{(m)},f(u)\subset
g(u)\Longleftrightarrow\forall u\in\widetilde{S}^{(m)},\widetilde{S}%
^{(n)}\setminus g(u)\subset\widetilde{S}^{(n)}\setminus
f(u)\Longleftrightarrow
\]%
\[
\Longleftrightarrow\forall u\in\widetilde{S}^{(m)},Cg(u)\subset
Cf(u)\Longleftrightarrow Cg\subset Cf
\]
\end{proof}

\begin{theorem}
$(Cf)^{\ast}=Cf^{\ast}$
\end{theorem}

\begin{proof}
For any $u\in\widetilde{S}^{(m)}$ we can write%
\[
(Cf)^{\ast}(u)=\{\overline{x}|x\in(Cf)(\overline{u})\}=\{\overline{x}%
|x\in\widetilde{S}^{(n)}\setminus f(\overline{u})\}=\{x|\overline{x}%
\in\widetilde{S}^{(n)}\setminus f(\overline{u})\}=
\]%
\[
=\{x|x\in\widetilde{S}^{(n)}\setminus\{z|\overline{z}\in f(\overline
{u})\}\}=\{x|x\in\widetilde{S}^{(n)}\setminus\{\overline{z}|z\in
f(\overline{u})\}\}=
\]%
\[
=\{x|x\in\widetilde{S}^{(n)}\setminus f^{\ast}(u)\}=(Cf^{\ast})(u)
\]
\end{proof}

\begin{theorem}
$(Cf)^{-1}=Cf^{-1}$
\end{theorem}

\begin{proof}
For all $x\in\widetilde{S}^{(n)}$ we have%
\[
(Cf)^{-1}(x)=\{u|u\in\widetilde{S}^{(m)},x\in Cf(u)\}=\{u|u\in\widetilde
{S}^{(m)},x\in\widetilde{S}^{(n)}\setminus f(u)\}=
\]%
\[
=\{u|u\in\widetilde{S}^{(m)}\setminus\{v|x\in f(v)\}\}=\{u|u\in\widetilde
{S}^{(m)}\setminus f^{-1}(x)\}=(Cf^{-1})(x)
\]
\end{proof}

\begin{theorem}
Let $f:\widetilde{S}^{(m)}\rightarrow P(\widetilde{S}^{(n)}),$ $f^{\prime
}:\widetilde{S}^{(m^{\prime})}\rightarrow P(\widetilde{S}^{(n^{\prime})})$ two
pseudo-systems. We have $Cf\times Cf^{\prime}\subset C(f\times f^{\prime})$.
\end{theorem}

\begin{proof}
$\forall(u,u^{\prime})\in\widetilde{S}^{(m+m^{\prime})},$%
\[
(Cf\times Cf^{\prime})(u,u^{\prime})=\{(x,x^{\prime})|x\in Cf(u),x^{\prime}\in
Cf^{\prime}(u^{\prime})\}=
\]%
\[
=\{(x,x^{\prime})|x\in\widetilde{S}^{(n)}\quad and\quad x\notin f(u)\quad
and\quad x^{\prime}\in\widetilde{S}^{(n^{\prime})}\quad and\quad x^{\prime
}\notin f^{\prime}(u^{\prime})\}=
\]%
\[
=\{(x,x^{\prime})|(x,x^{\prime})\in\widetilde{S}^{(n+n^{\prime})}\quad
and\quad x\notin f(u)\quad and\quad x^{\prime}\notin f^{\prime}(u^{\prime
})\}\subset
\]%
\[
\subset\{(x,x^{\prime})|(x,x^{\prime})\in\widetilde{S}^{(n+n^{\prime})}\quad
and\quad(x\notin f(u)\quad or\quad x^{\prime}\notin f^{\prime}(u^{\prime
}))\}=
\]%
\[
=\{(x,x^{\prime})|(x,x^{\prime})\in\widetilde{S}^{(n+n^{\prime})}\quad
and\quad(x,x^{\prime})\notin(f\times f^{\prime})(u,u^{\prime})\}=C(f\times
f^{\prime})(u,u^{\prime})
\]
\end{proof}

\begin{theorem}
For $f:\widetilde{S}^{(m)}\rightarrow P(\widetilde{S}^{(n)}),$ $f^{\prime
}:\widetilde{S}^{(m)}\rightarrow P(\widetilde{S}^{(n^{\prime})})$ we can write
$(Cf,Cf^{\prime})\subset C(f,f^{\prime})$.
\end{theorem}

\section{Intersection and reunion}

\begin{definition}
Let the pseudo-systems $f,g:\widetilde{S}^{(m)}\rightarrow P(\widetilde
{S}^{(n)})$. The pseudo-systems $f\cap g,f\cup g:\widetilde{S}^{(m)}%
\rightarrow P(\widetilde{S}^{(n)})$ are defined by%
\[
\forall u\in\widetilde{S}^{(m)},(f\cap g)(u)=f(u)\cap g(u)
\]%
\[
\forall u\in\widetilde{S}^{(m)},(f\cup g)(u)=f(u)\cup g(u)
\]
\end{definition}

\begin{remark}
The intersection of the pseudo-systems represents the gain of information (of
precission) in the modeling of a circuit by considering the validity of two
models at the same time. The reunion of the pseudo-systems is the dual concept
representing the loss of information (of precission) in modeling as a result
of considering the validity of one of two models.

The set of the $\widetilde{S}^{(m)}\rightarrow P(\widetilde{S}^{(n)})$
pseudo-systems is a Boole algebra relative to $C,\cap,\cup$. The zero and the
one of this Boole algebra are the null and the total pseudo-systems.
\end{remark}

\begin{theorem}
Let the pseudo-systems $f$ and $g$. If $f$ has initial states (race-free
initial states, constant initial state), then $f\cap g$ has initial states
(race-free initial states, constant initial state).
\end{theorem}

\begin{proof}
$f\cap g\subset f$ and the statement of the theorem follows from Theorem 7.3.
\end{proof}

\begin{theorem}
If $f$ has final states (race-free final states, constant final state), then
$f\cap g$ has final states (race-free final states, constant final state).
\end{theorem}

\begin{theorem}
If the pseudo-systems $f,g$ have initial states (a common constant initial
state), then $f\cup g$ has initial states (constant initial state).
\end{theorem}

\begin{theorem}
If $f,g$ have final states (a common constant final state), then $f\cup g$ has
final states (constant final state).
\end{theorem}

\begin{remark}
The statements of the previous two theorems are not true in general for the
pseudo-systems $f,g$ with race-free initial states and for the pseudo-systems
$f,g$ with constant initial states, because it is possible that the two
partial functions $\widetilde{S}^{(m)}\rightarrow\mathbf{B}^{n}$ from Remark
4.1 corresponding to $f$ and $g$ differ, respectively that the two constant
initial states corresponding to $f$ and $g$ differ. Similar reasoning for the
final states. Such 'disappearances of the middle statement', could be the
race-free statement about the initial/final states, could be the boundness
statement about the initial/final time, have already occurred (for different
reasons) at theorems 12.2,...,12.5.
\end{remark}

\begin{theorem}
If $f$ has unbounded initial time (bounded initial time, fix initial time),
then $f\cap g$ has unbounded initial time (bounded initial time, fix initial time).
\end{theorem}

\begin{proof}
$f\cap g\subset f$ and the results follow from Theorem 7.5.
\end{proof}

\begin{theorem}
If $f$ has unbounded final time (bounded final time, fix final time), then
$f\cap g$ has unbounded final time (bounded final time, fix final time).
\end{theorem}

\begin{theorem}
If $f,g$ have unbounded initial time (bounded initial time, fix initial time),
then $f\cup g$ has unbounded initial time (bounded initial time, fix initial time).
\end{theorem}

\begin{theorem}
If $f,g$ have unbounded final time (bounded final time, fix final time), then
$f\cup g$ has unbounded final time (bounded final time, fix final time).
\end{theorem}

\begin{proof}
We suppose for example that $f$ and $g$ satisfy (\ref{e3.5}), i.e. they have
bounded final time. For some arbitrary $u\in\widetilde{S}^{(m)}$, let
$t_{f},t_{f}^{\prime}$ the final time instants of $f$, respectively of $g$.
Then (\ref{e3.5}) is satisfied by $f\cup g$ because we can choose for $u$ the
final time instant $t_{f}^{"}\geq\max(t_{f},t_{f}^{\prime})$.
\end{proof}

\begin{theorem}
We suppose that $f,g$ have initial states. We have $(\phi\cap\gamma)_{0}%
,(\phi\cup\gamma)_{0}:\widetilde{S}^{(m)}\rightarrow P(\mathbf{B}^{n}),$%
\[
\forall u\in\widetilde{S}^{(m)},(\phi\cap\gamma)_{0}(u)=\phi_{0}(u)\cap
\gamma_{0}(u)
\]%
\[
\forall u\in\widetilde{S}^{(m)},(\phi\cup\gamma)_{0}(u)=\phi_{0}(u)\cup
\gamma_{0}(u)
\]%
\[
(\Theta\cap\Gamma)_{0}=\underset{u\in\widetilde{S}^{(m)}}{\bigcup}\phi
_{0}(u)\cap\gamma_{0}(u)
\]%
\[
(\Theta\cup\Gamma)_{0}=\underset{u\in\widetilde{S}^{(m)}}{\bigcup}\phi
_{0}(u)\cup\gamma_{0}(u)
\]
We have noted with $\phi_{0},\gamma_{0},(\phi\cap\gamma)_{0},(\phi\cup
\gamma)_{0}$ the initial state functions of $f,g,f\cap g,f\cup g$ and with
$(\Theta\cap\Gamma)_{0},(\Theta\cup\Gamma)_{0}$ the sets of initial states of
$f\cap g,f\cup g$.
\end{theorem}

\begin{proof}
$f\cup g$ has initial states from Theorem 14.5, thus $(\phi\cup\gamma)_{0}$
and $(\Theta\cup\Gamma)_{0}$ exist. We can write that $\forall u\in
\widetilde{S}^{(m)},$%
\[
(\phi\cup\gamma)_{0}(u)=\{x(-\infty+0)|x\in(f\cup g)(u)\}=\{x(-\infty+0)|x\in
f(u)\cup g(u)\}=
\]%
\[
=\{x(-\infty+0)|x\in f(u)\}\cup\{x(-\infty+0)|x\in g(u)\}=\phi_{0}%
(u)\cup\gamma_{0}(u)
\]
\end{proof}

\begin{theorem}
If $f,g$ have final states, then we have $(\phi\cap\gamma)_{f},(\phi\cup
\gamma)_{f}:\widetilde{S}^{(m)}\rightarrow P(\mathbf{B}^{n}),$%
\[
\forall u\in\widetilde{S}^{(m)},(\phi\cap\gamma)_{f}(u)=\phi_{f}(u)\cap
\gamma_{f}(u)
\]%
\[
\forall u\in\widetilde{S}^{(m)},(\phi\cup\gamma)_{f}(u)=\phi_{f}(u)\cup
\gamma_{f}(u)
\]%
\[
(\Theta\cap\Gamma)_{f}=\underset{u\in\widetilde{S}^{(m)}}{\bigcup}\phi
_{f}(u)\cap\gamma_{f}(u)
\]%
\[
(\Theta\cup\Gamma)_{f}=\underset{u\in\widetilde{S}^{(m)}}{\bigcup}\phi
_{f}(u)\cup\gamma_{f}(u)
\]
The notations are obvious and similar with those from the previous theorem.
\end{theorem}

\begin{theorem}
We have%
\[
(f\cap g)^{\ast}=f^{\ast}\cap g^{\ast}%
\]%
\[
(f\cup g)^{\ast}=f^{\ast}\cup g^{\ast}%
\]
\end{theorem}

\begin{proof}
$\forall u\in\widetilde{S}^{(m)},$%
\[
(f\cup g)^{\ast}(u)=\{\overline{x}|x\in(f\cup g)(\overline{u})\}=\{\overline
{x}|x\in f(\overline{u})\cup g(\overline{u})\}=
\]%
\[
=\{\overline{x}|x\in f(\overline{u})\}\cup\{\overline{x}|x\in g(\overline
{u})\}=f^{\ast}(u)\cup g^{\ast}(u)=(f^{\ast}\cup g^{\ast})(u)
\]
\end{proof}

\begin{theorem}
The next formulas of inversion take place:%
\[
(f\cap g)^{-1}=f^{-1}\cap g^{-1}%
\]%
\[
(f\cup g)^{-1}=f^{-1}\cup g^{-1}%
\]
\end{theorem}

\begin{proof}
$\forall x\in\widetilde{S}^{(n)},$%
\[
(f\cap g)^{-1}(x)=\{u|x\in(f\cap g)(u)\}=\{u|x\in f(u)\cap g(u)\}=
\]%
\[
=\{u|x\in f(u)\}\cap\{u|x\in g(u)\}=f^{-1}(x)\cap g^{-1}(x)=(f^{-1}\cap
g^{-1})(x)
\]
\end{proof}

\begin{theorem}
Let the pseudo-systems $f,g:\widetilde{S}^{(m)}\rightarrow P(\widetilde
{S}^{(n)}),f^{\prime},g^{\prime}:\widetilde{S}^{(m^{\prime})}\rightarrow
P(\widetilde{S}^{(n^{\prime})})$. The next statements are true:%
\[
(f\cap g)\times(f^{\prime}\cap g^{\prime})=(f\times f^{\prime})\cap(g\times
g^{\prime})
\]%
\[
(f\cup g)\times(f^{\prime}\cup g^{\prime})=(f\times f^{\prime})\cup(g\times
g^{\prime})
\]
\end{theorem}

\begin{proof}
We have $\forall(u,u^{\prime})\in\widetilde{S}^{(m+m^{\prime})}$:%
\[
((f\cap g)\times(f^{\prime}\cap g^{\prime}))(u,u^{\prime})=\{(x,x^{\prime
})|x\in(f\cap g)(u)\quad and\quad x^{\prime}\in(f^{\prime}\cap g^{\prime
})(u^{\prime})\}=
\]%
\[
=\{(x,x^{\prime})|x\in f(u)\cap g(u)\quad and\quad x^{\prime}\in f^{\prime
}(u^{\prime})\cap g^{\prime}(u^{\prime})\}=
\]%
\[
=\{(x,x^{\prime})|x\in f(u)\quad and\quad x\in g(u)\quad and\quad x^{\prime
}\in f^{\prime}(u^{\prime})\quad and\quad x^{\prime}\in g^{\prime}(u^{\prime
})\}=
\]%
\[
=\{(x,x^{\prime})|x\in f(u)\quad and\quad x^{\prime}\in f^{\prime}(u^{\prime
})\quad and\quad x\in g(u)\quad and\quad x^{\prime}\in g^{\prime}(u^{\prime
})\}=
\]%
\[
=\{(x,x^{\prime})|x\in f(u)\quad and\quad x^{\prime}\in f^{\prime}(u^{\prime
})\}\cap\{(x,x^{\prime})|x\in g(u)\quad and\quad x^{\prime}\in g^{\prime
}(u^{\prime})\}=
\]%
\[
=(f\times f^{\prime})(u,u^{\prime})\cap(g\times g^{\prime})(u,u^{\prime
})=((f\times f^{\prime})\cap(g\times g^{\prime}))(u,u^{\prime})
\]
\end{proof}

\begin{theorem}
We consider the pseudo-systems $f,g:\widetilde{S}^{(m)}\rightarrow
P(\widetilde{S}^{(n)}),f^{\prime},g^{\prime}:\widetilde{S}^{(m)}\rightarrow
P(\widetilde{S}^{(n^{\prime})})$. We have:%
\[
(f\cap g,f^{\prime}\cap g^{\prime})=(f,f^{\prime})\cap(g,g^{\prime})
\]
\[
(f\cup g,f^{\prime}\cup g^{\prime})=(f,f^{\prime})\cup(g,g^{\prime})
\]
\end{theorem}

\begin{theorem}
For the pseudo-systems $f,g$ and $h,h_{1}:\widetilde{S}^{(n)}\rightarrow
P(\widetilde{S}^{(p)})$ we have:%
\[
h\circ(f\cap g)\subset(h\circ f)\cap(h\circ g)
\]%
\[
h\circ(f\cup g)\subset(h\circ f)\cup(h\circ g)
\]%
\[
(h\cap h_{1})\circ f\subset(h\circ f)\cap(h_{1}\circ f)
\]%
\[
(h\cup h_{1})\circ f\subset(h\circ f)\cup(h_{1}\circ f)
\]
\end{theorem}

\begin{proof}
$\forall u\in\widetilde{S}^{(m)},$%
\[
(h\circ(f\cap g))(u)=\{y|\exists x,x\in(f\cap g)(u),y\in h(x)\}=\{y|\exists
x,x\in f(u)\cap g(u),y\in h(x)\}=
\]%
\[
=\{y|\exists x,x\in f(u)\quad and\quad x\in g(u)\quad and\quad y\in h(x)\}=
\]%
\[
=\{y|\exists x,x\in f(u)\quad and\quad y\in h(x)\quad and\quad x\in g(u)\quad
and\quad y\in h(x)\}\subset
\]%
\[
\subset\{y|\exists x,x\in f(u)\quad and\quad y\in h(x)\quad and\quad\exists
z,z\in g(u)\quad and\quad y\in h(z)\}=
\]%
\[
=\{y|\exists x,x\in f(u)\quad and\quad y\in h(x)\}\cap\{y|\exists z,z\in
g(u)\quad and\quad y\in h(z)\}=
\]%
\[
=(h\circ f)(u)\cap(h\circ g)(u)=((h\circ f)\cap(h\circ g))(u)
\]
$\forall u\in\widetilde{S}^{(m)},$%
\[
((h\cup h_{1})\circ f)(u)=\{y|\exists x,x\in f(u),y\in(h\cup h_{1})(x)\}=
\]%
\[
=\{y|\exists x,x\in f(u)\quad and\quad y\in h(x)\cup h_{1}(x)\}=
\]%
\[
=\{y|\exists x,x\in f(u)\quad and\quad y\in h(x)\quad or\quad x\in f(u)\quad
and\quad y\in h_{1}(x)\}\subset
\]%
\[
\subset\{y|\exists x,x\in f(u)\quad and\quad y\in h(x)\quad or\quad\exists
z,z\in f(u)\quad and\quad y\in h_{1}(z)\}=
\]%
\[
=\{y|\exists x,x\in f(u)\quad and\quad y\in h(x)\}\cup\{y|\exists z,z\in
f(u)\quad and\quad y\in h_{1}(z)\}=
\]%
\[
=(h\circ f)(u)\cup(h_{1}\circ f)(u)=((h\circ f)\cup(h_{1}\circ f))(u)
\]
\end{proof}

\section{Systems}

\begin{definition}
Let the pseudo-system $f:\widetilde{S}^{(m)}\rightarrow P(\widetilde{S}%
^{(n)})$. The set $U_{f}$ of the admissible inputs defined by%
\[
U_{f}=\{u|u\in\widetilde{S}^{(m)},f(u)\neq\emptyset\}
\]
is also called the support (set) of $f$.
\end{definition}

\begin{definition}
The (asynchronous) pseudo-system $f$ is called (asynchronous) system if

a) $U_{f}\neq\emptyset$

b) $U_{f}\subset S^{(m)}$

c) $\forall u\in U_{f},f(u)\subset S^{(n)}$.
\end{definition}

\begin{remark}
We shall identify the system $f$ with the function $f_{1}:U\rightarrow
P^{\ast}(S^{(n)})$, where $U=U_{f}$, that is defined by $\forall u\in
U,f_{1}(u)=f(u)$. We shall also identify the initial state function $\phi
_{0}:\widetilde{S}^{(m)}\rightarrow P(\mathbf{B}^{n})$ with the function
$\phi_{10}:U\rightarrow P^{\ast}(\mathbf{B}^{n})$ defined by $\forall u\in
U,\phi_{10}(u)=\phi_{0}(u)$.
\end{remark}

\begin{notation}
The systems are noted sometimes with $f:U\rightarrow P^{\ast}(S^{(n)})$, where
$U\subset S^{(m)}$ is non-empty. If $\forall u\in U,f(u)$ has a single
element, then we have the usual notation $f:U\rightarrow S^{(n)}$ of the
uni-valued functions. Similarly, their initial state functions are noted
sometimes with $\phi_{0}:U\rightarrow P^{\ast}(\mathbf{B}^{n})$ or with
$\phi_{0}:U\rightarrow\mathbf{B}^{n}$ when $\forall u\in U,x(-\infty+0)$ is unique.
\end{notation}

\begin{remark}
The systems are those non-null pseudo-systems $f$ for which the admissible
inputs and the possible states are signals (resulting that $f$ has initial
states). The concept creates an asymmetry between the initial states and the
final states because:

- it is natural that the inputs be considered commands, a deliberate manner of
acting on the circuit modeled by $f$ with the purpose of producing a certain
effect. But this is made after choosing an initial time instant $t_{0}$ from
which we order our actions in the increasing sense of the time axis (not in
both senses)

- it is natural that we associate to the request $U\subset S^{(m)}$ a request
(Definition 15.2 c)) that is dual to stability: the system orders its
reactions in the increasing sense of the time axis (not in both senses).
\end{remark}

\begin{example}
The fact that (\ref{e6.1}) defines a $S\rightarrow P^{\ast}(S)$ system is
obvious if we observe that%
\[
\forall u\in S,\forall\tau\in(0,d],\underset{\xi\in\lbrack t-d,t)}{\bigcap
}u(\xi)\leq u(t-\tau)\leq\underset{\xi\in\lbrack t-d,t)}{\bigcup}u(\xi)
\]
thus $x(t)=u(t-\tau)$, which is a signal, satisfies it whenever $\tau\in
(0,d]$. This system is called the symmetrical upper bounded, lower unbounded delay.
\end{example}

\begin{notation}
Let $f:\widetilde{S}^{(m)}\rightarrow P(\widetilde{S}^{(n)})$ a pseudo-system
with the property that%
\begin{equation}
\exists u\in S^{(m)},f(u)\cap S^{(n)}\neq\emptyset\label{e6.2}%
\end{equation}
We note with $[f]:U\rightarrow P^{\ast}(S^{(n)})$ the function that is defined
by%
\begin{equation}
U=\{u|u\in S^{(m)},f(u)\cap S^{(n)}\neq\emptyset\}\label{e6.3}%
\end{equation}%
\begin{equation}
\forall u\in U,[f](u)=f(u)\cap S^{(n)}\label{e6.4}%
\end{equation}
\end{notation}

\begin{theorem}
a) $[f]$ is a system

b) $[f]\subset f$

c) Let $g:\widetilde{S}^{(m)}\rightarrow P(\widetilde{S}^{(n)})$ a system so
that $g\subset f$. Then $g\subset\lbrack f]$, i.e. $[f]$ is the greatest
system that is included in $f$.
\end{theorem}

\begin{proof}
a) $U\neq\emptyset$ follows from (\ref{e6.2}) and (\ref{e6.3}), $U\subset
S^{(m)}$ is a consequence of (\ref{e6.3}) and $\forall u\in U,[f](u)\subset
S^{(n)}$ results from (\ref{e6.4}), thus $[f]$ is a system.

b) From (\ref{e6.4})

c) Let $g:\widetilde{S}^{(m)}\rightarrow P(\widetilde{S}^{(n)})$ a system so
that $\forall u\in\widetilde{S}^{(m)},g(u)\subset f(u)$, from where $\forall
u\in\widetilde{S}^{(m)},g(u)=g(u)\cap S^{(n)}\subset f(u)\cap S^{(n)}=[f](u)$
\end{proof}

\begin{definition}
When the pseudo-system $f$ satisfies the property (\ref{e6.2}), $[f]$ is
called the system that is induced by $f$.
\end{definition}

\begin{theorem}
The pseudo-system $f$ is a system if and only if $f=[f]$.
\end{theorem}

\begin{proof}
$\Longleftarrow$ is obvious, since $[f]$ is a system.

$\Longrightarrow$ Admissible inputs exist and let $u$ such an input. Because
$f$ is a system, $u$ is signal. From $f(u)\subset S^{(n)}$, we have that
$f(u)=f(u)\cap S^{(n)}$ and as $u$ was arbitrarily chosen we infer that
$f=[f]$.
\end{proof}

\begin{theorem}
For any system $f$, the initial state function $\phi_{0}$ and the set of the
initial states $\Theta_{0}$ exist.
\end{theorem}

\begin{theorem}
Let the systems $f:U\rightarrow P^{\ast}(S^{(n)}),$ $g:V\rightarrow P^{\ast
}(S^{(n)}),$ $U,V\in P^{\ast}(S^{(m)})$. We have%
\[
f\subset g\Longleftrightarrow U\subset V\quad and\quad\forall u\in
U,f(u)\subset g(u)
\]
\end{theorem}

\begin{proof}
$f\subset g\Longrightarrow U\subset V\quad and\quad\forall u\in U,f(u)\subset g(u)$

Each of the suppositions $U\setminus V\neq\emptyset$ and respectively $\exists
u\in U,\exists x\in f(u)$ so that $x\in S^{(n)}\setminus g(u)$ gives a
contradiction with the hypothesis $f\subset g$

$U\subset V\quad and\quad\forall u\in U,f(u)\subset g(u)\Longrightarrow
f\subset g$

The implication is obvious.
\end{proof}

\begin{theorem}
If $f$ is a system, then its dual $f^{\ast}$ is a system too.
\end{theorem}

\begin{theorem}
For the system $f$, the function $f^{-1}:X\rightarrow P^{\ast}(S^{(m)})$ given
by%
\[
X=\{x|\exists u\in U,x\in f(u)\}
\]%
\[
\forall x\in X,f^{-1}(x)=\{u|u\in U,x\in f(u)\}
\]
is a system that coincides with the inverse of $f$ (as pseudo-system).
\end{theorem}

\begin{proof}
From the hypothesis, the support $U$ of $f$ is non-empty so that we have
$X\neq\emptyset$. The fact that $U\subset S^{(m)}$ implies $\forall x\in
X,f^{-1}(x)\subset S^{(m)}$ and $\forall u\in U,f(u)\subset S^{(n)}$ gives
$X\subset S^{(n)}$, thus $f^{-1}$ is a system. $f^{-1}$ obviously coincides
with the inverse of $f$ as pseudo-system.
\end{proof}

\begin{theorem}
The direct product of two systems is a system.
\end{theorem}

\begin{proof}
We consider the systems $f:U\rightarrow P^{\ast}(S^{(n)}),U\in P^{\ast
}(S^{(m)})$ and $f^{\prime}:U^{\prime}\rightarrow P^{\ast}(S^{(n^{\prime}%
)}),U^{\prime}\in P^{\ast}(S^{(m^{\prime})})$. We remark that $U\times
U^{\prime}\in P^{\ast}(S^{(m+m^{\prime})})$ and $\forall(u,u^{\prime})\in
U\times U^{\prime},(f\times f^{\prime})(u,u^{\prime})\in P^{\ast
}(S^{(n+n^{\prime})})$, thus $f\times f^{\prime}$ is a system.
\end{proof}

\begin{theorem}
Let the systems $f:U\rightarrow P^{\ast}(S^{(n)}),$ $f^{\prime}:U^{\prime
}\rightarrow P^{\ast}(S^{(n^{\prime})}),$ $U,U^{\prime}\in P^{\ast}(S^{(m)})$.
Their parallel connection is a system if and only if $U\cap U^{\prime}%
\neq\emptyset$. In this case we have $(f,f^{\prime}):U\cap U^{\prime
}\rightarrow P^{\ast}(S^{(n+n^{\prime})}),$%
\[
\forall u\in U\cap U^{\prime},(f,f^{\prime})(u)=\{(x,x^{\prime})|(x,x^{\prime
})\in S^{(n+n^{\prime})},x\in f(u),x^{\prime}\in f^{\prime}(u)\}
\]
\end{theorem}

\begin{theorem}
We consider the systems $f:U\rightarrow P^{\ast}(S^{(n)}),$ $U\in P^{\ast
}(S^{(m)})$ and $h:X\rightarrow P^{\ast}(S^{(p)}),$ $X\in P^{\ast}(S^{(n)})$.
Their serial connection is a system if and only if $\exists u\in U,f(u)\cap
X\neq\emptyset$. In the case that this condition is fulfilled, we note
\[
W=\{u|u\in U,f(u)\cap X\neq\emptyset\}
\]
and we have $h\circ f:W\rightarrow P^{\ast}(S^{(p)})$,%
\[
\forall u\in W,(h\circ f)(u)=\{y|\exists x\in f(u)\cap X,y\in h(x)\}
\]
\end{theorem}

\begin{remark}
Given the system $f$, its complement $Cf$ is a pseudo-system since
\[
\forall u\in\widetilde{S}^{(m)},Cf(u)=\widetilde{S}^{(n)}\setminus
f(u)\supset\widetilde{S}^{(n)}\setminus S^{(n)}%
\]
If $f$ is a pseudo-system, then $Cf$ can be a system or a pseudo-system.
\end{remark}

\begin{theorem}
We consider the systems $f:U\rightarrow P^{\ast}(S^{(n)}),$ $g:V\rightarrow
P^{\ast}(S^{(n)}),$ $U,V\in P^{\ast}(S^{(m)})$. Their intersection is a system
if and only if
\[
\exists u\in U\cap V,f(u)\cap g(u)\neq\emptyset
\]
In the case when this condition is fulfilled, we have $f\cap g:W\rightarrow
P^{\ast}(S^{(n)})$,%
\[
W=\{u|u\in U\cap V,f(u)\cap g(u)\neq\emptyset\}
\]%
\[
\forall u\in W,(f\cap g)(u)=f(u)\cap g(u)
\]
\end{theorem}

\begin{proof}
$W\neq\emptyset$ is the support set of $f\cap g$; we obtain $W\subset U\cap
V\subset S^{(m)}$ and on the other hand we get $\forall u\in W,(f\cap
g)(u)\subset f(u)\subset S^{(n)}$, thus $f\cap g$ is a system.
\end{proof}

\begin{theorem}
The reunion of the systems $f$ and $g$ is the system $f\cup g:U\cup
V\rightarrow P^{\ast}(S^{(n)})$ that is defined in the next manner%
\[
\forall u\in U\cup V,(f\cup g)(u)=f(u)\cup g(u)
\]
\end{theorem}

\begin{proof}
$U\neq\emptyset$ and $V\neq\emptyset$ imply $U\cup V\neq\emptyset$; $U\subset
S^{(m)}$ and $V\subset S^{(m)}$ imply $U\cup V\subset S^{(m)}$; and $\forall
u\in U,f(u)\subset S^{(n)}$, $\forall u\in V,g(u)\subset S^{(n)}$ imply
$\forall u\in U\cup V,f(u)\cup g(u)\subset S^{(n)}$, thus $f\cup g$ is a system
\end{proof}

\end{document}